\documentclass[journal, final]{IEEEtran}
\usepackage{latexsym}
\usepackage{graphicx}  % Written by David Carlisle and Sebastian Rahtz
\usepackage{psfrag}    % Written by Craig Barratt, Michael C. Grant,
\usepackage{subfigure} % Written by Steven Douglas Cochran
\usepackage{url}       % Written by Donald Arseneau
\usepackage{overpic}       % Written by Donald Arseneau
\usepackage{amsmath}   % From the American Mathematical Society
\usepackage{amssymb}   % From the American Mathematical Society

\newtheorem{theo}{Theorem}

\newcommand{\D}{\displaystyle}

\begin{document}

\title{Gaussian Z-Interference Channel with a Relay Link:
Achievability Region and Asymptotic Sum Capacity
\thanks{
Manuscript submitted to the {\it IEEE Transactions on Information
Theory} on Sept 3, 2008, resubmitted on June 10, 2010 and revised on
June 8, 2011.  The material in this paper has been presented
in part at the IEEE International Symposium on Information Theory and
its Applications (ISITA), Auckland, New Zealand, December 2008, and in
part at the IEEE Information Theory and Applications (ITA) Workshop,
San Diego, CA, February 2009.  The authors are with the Electrical and
Computer Engineering Department, University of Toronto, 10 King's
College Road, Toronto, Ontario M5S 3G4, Canada (email:
zhoulei@comm.utoronto.ca; weiyu@comm.utoronto.ca).  This work was
supported in part by the Natural Science and Engineering Research
Council (NSERC) of Canada under the Canada Research Chairs program,
and in part by the Ontario Early Researcher Awards program.  Kindly
address correspondence to Lei Zhou (zhoulei@comm.utoronto.ca).
}
}

\author{ Lei Zhou, {\it Student Member, IEEE} and
	Wei Yu, {\it Senior Member, IEEE}}

\markboth{IEEE TRANSACTIONS ON INFORMATION THEORY, VOL. 58, NO. 4, APRIL 2012}
{Zhou and Yu: Gaussian Z-Interference Channel with a Relay Link:
Achievability Region and Asymptotic Sum Capacity
}

\maketitle

\begin{abstract}
This paper studies a Gaussian Z-interference channel with a
rate-limited digital relay link from one receiver to another.
Achievable rate regions are derived based on a combination of
Han-Kobayashi common-private power splitting technique and
either a compress-and-forward relay strategy or
a decode-and-forward strategy for interference subtraction at the other end.  For the Gaussian Z-interference
channel with a digital link from the interference-free receiver to the
interfered receiver, the capacity region is established in the strong
interference regime; an achievable rate region is established in the
weak interference regime. In the weak interference regime, the
decode-and-forward strategy is shown to be asymptotically sum-capacity
achieving in the high signal-to-noise ratio and high
interference-to-noise ratio limit.  In this case, each relay bit
asymptotically improves the sum capacity by one bit.  For the Gaussian
Z-interference channel with a digital link from the interfered
receiver to the interference-free receiver, the capacity region is
established in the strong interference regime; achievable rate regions
are established in the moderately strong and weak interference
regimes.  In addition, the asymptotic sum capacity is established
in the limit of large relay link rate. In this case, the sum capacity
improvement due to the digital link is bounded by half a bit when the
interference link is weaker than a certain threshold, but the sum
capacity improvement becomes unbounded when the interference link
is strong.
\end{abstract}


\section{Introduction}

The classic interference channel models a communication situation
in which two transmitters communicate with their respective intended
receivers while mutually interfering with each other. The
interference channel is of fundamental importance for communication
system design, because many practical systems are designed to operate
in the interference-limited regime. The largest known achievability
region for the interference channel is due to Han and Kobayashi
\cite{HK1981}, where a common-private power splitting
technique is used to partially decode and subtract the interfering
signal. The Han-Kobayashi scheme has been shown to be
capacity achieving in a very weak interference regime \cite{VVV,
Khandani08, Biao}  and to be within one bit of the capacity region
in general \cite{Tse2007}.

This paper considers a communication model in which the classic
interference channel is augmented by a noiseless relay link between
the two receivers.  We are motivated to study such a relay-interference channel because in practical wireless cellular systems,
the uplink receivers at the base-stations are connected via backhaul links and
the downlink receivers may also be capable of establishing an
independent communication link for the purpose of interference
mitigation.

\begin{figure*}
\centering \psfrag{X1}{$X_1$} \psfrag{X2}{$X_2$} \psfrag{Z1}{$Z_1$}
\psfrag{Z2}{$Z_2$} \psfrag{Y1}{$Y_1$} \psfrag{Y2}{$Y_2$}
\psfrag{h11}{$h_{11}$}  \psfrag{h22}{$h_{22}$}
\psfrag{h21}{$h_{21}$} \psfrag{h12}{$h_{12}$} \psfrag{R21}{$R_{21}$}
\psfrag{R12}{$R_{12}$} \psfrag{R0}{$R_0$}
\psfrag{(a) Type I}{(a) Type I}
\psfrag{(b) Type II}{(b) Type II}
\includegraphics[width=1.950in]{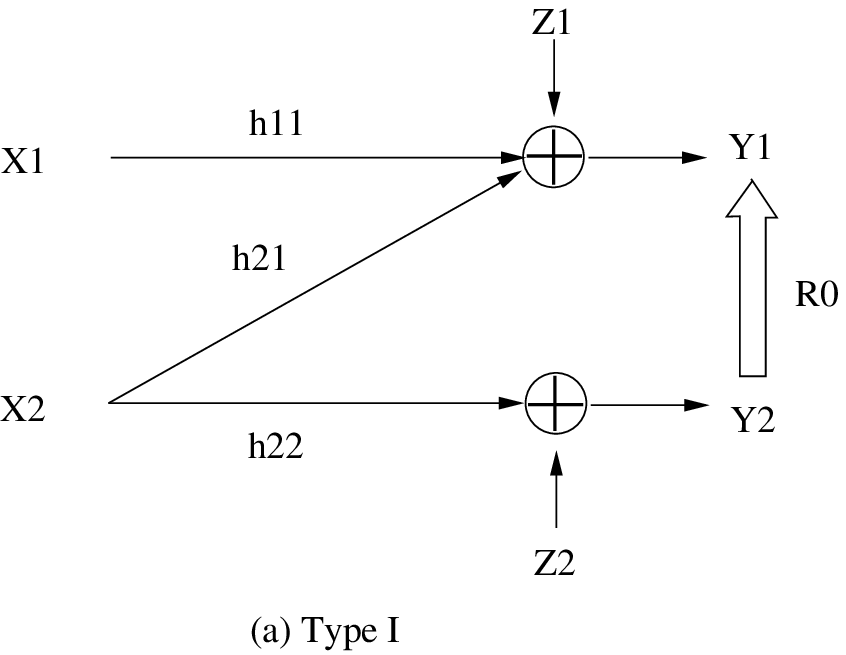}
\hspace{1in}
\includegraphics[width=1.950in]{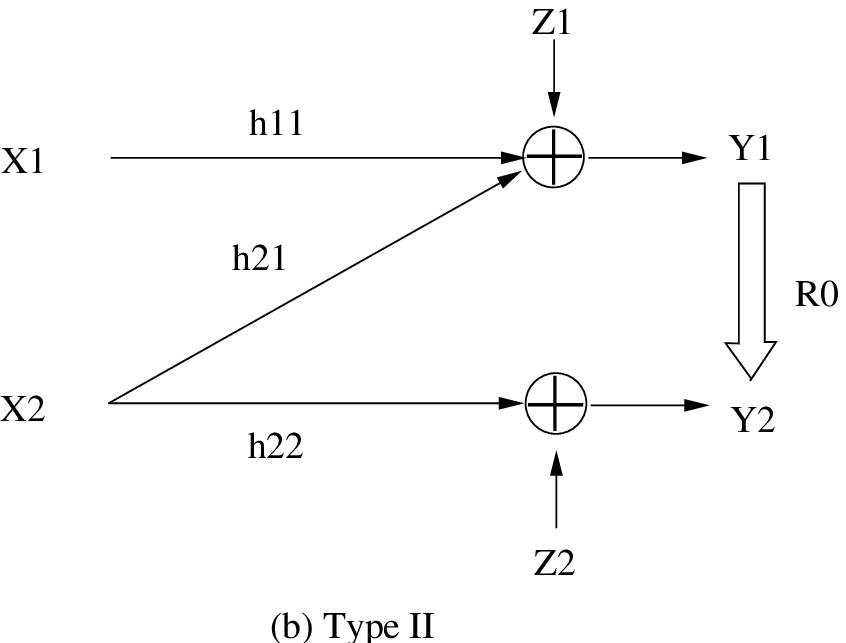}
\caption{Gaussian Z-interference channel with a relay link:
(a) Type I; (b) Type II.} \label{Gaussian_Z_introduction}
\end{figure*}

This paper explores the use of relay techniques for interference
mitigation.  We focus on the simplest interference channel model,
the Gaussian Z-interference channel (also known as the one-sided
interference channel), in which one of the receivers gets an
interference-free signal, the other receiver gets a combination of the
intended and the interfering signals, and the channel is equipped with
a noiseless link of fixed capacity from one receiver to the other.
The Z-interference channel is of practical interest because it models
a two-cell cellular network with one user located at the cell edge
and another user at the cell center. (The
cell-edge user is sometimes referred to as in a soft-handoff mode
\cite{Somekh_softhandoff}.) Depending on the direction of the noiseless link, the
proposed model is named the Type I or the Type II Gaussian
Z-relay-interference channel in this paper as shown in
Fig.~\ref{Gaussian_Z_introduction}.

The Type I Gaussian Z-relay-interference channel has a digital relay
link of finite capacity from the interference-free receiver to the
interfered receiver. Our main coding strategy for the Type I channel
is a decode-and-forward strategy, in which the relay link
forwards part of the interference to the interfered receiver using
a binning technique for interference subtraction.
%The idea is to exploit the
%fact that interference consists of structured codewords, which can be
%described efficiently using a binning technique.  This extra piece of
%binning information enhances the interference subtraction capability
%at the interfered receiver, thereby enlarging the overall achievable
%region for the interference channel.
This paper shows that decode-and-forward is capacity achieving
for the Type I channel in the strong interference regime, and is
asymptotically sum-capacity achieving in the weak interference regime.
In addition, in the weak interference regime, every bit of relay link
rate increases the sum rate by one bit in the high signal-to-noise
ratio (SNR) and high interference-to-noise ratio (INR) limit.

The Type II Gaussian Z-relay-interference channel differs from the
Type I channel in that the direction of the digital link goes
from the interfered receiver to the interference-free receiver.
%In this channel model, receiver $1$ now observes a noisy
%version of user 2's transmit signal and acts as a relay for user 2.
Our main coding strategy for the Type II channel is based on
a combination of two relaying strategies: decode-and-forward and
compress-and-forward. In the proposed scheme, the interfered receiver,
which decodes the common message and observes a noisy version of
the neighbor's private message, describes the common message with
a bin index and describes the neighbor's private message using
a quantization scheme. It is shown that, in the strong interference
regime, a special form of the proposed relaying scheme, which uses
decode-and-forward only, is capacity achieving. In the weak interference regime,
the proposed scheme reduces to pure compress-and-forward.
Further, when the interference link is weaker
than a certain threshold, the sum-capacity gain due to the digital link
for the Type II channel is upper bounded by half a bit. This is in
contrast to the Type I channel, in which each relay bit can be worth
up to one bit in sum capacity.
%Thus, it is much more beneficial for the "rich" to help the "poor",
%rather than the other way around!

\subsection{Related Work}

The Gaussian Z-interference channel has been extensively studied in
the literature. It is one of the few examples of an interference channel
(besides the strong interference case \cite{HK1981, Carleial, Sato}
and the very weak interference case \cite{VVV, Khandani08, Biao}) for
which the sum capacity has been established. The sum capacity of the
Gaussian Z-interference channel in the weak interference regime is
achieved with both transmitters using Gaussian codebooks and with the
interfered receiver treating the interference as noise
\cite{Tse2007, Sason2004}.

The fundamental decode-and-forward and compress-and-forward strategies
for the relay channel are due to the classic work of Cover and El Gamal
\cite{Cover1979}. Our study of the interference channel with a relay
link is motivated by the more recent capacity results for a class of
deterministic relay channels investigated by Kim \cite{Kim2008} and a
class of modulo-sum relay channels investigated by Aleksic et al.
\cite{marko}, where the relay observes the noise in the direct
channel. The situation investigated in \cite{Kim2008,marko} is similar
to the Type I Gaussian Z-relay-interference channel, where the
interference-free receiver observes a noisy version of the
interference at the interfered receiver and helps the interfered
receiver by describing the interference through a noiseless relay
link.

The channel model studied in the paper is related to the work of Sahin et al. \cite{Sahin_Erkip_1, Sahin_Erkip_2, Sahin_outofband_relay_ic}, Mari\'c et al.
\cite{Maric_Dabora}, Dabora et al. \cite{Dabora_Maric}, and Tian and Yener \cite{Tian_outofband_relay_ic}, where the
achievable rate regions and the relay strategies are studied for an interference channel with an additional relay node, and where the relay observes the transmitted signals from the inputs and contributes to the outputs of both channels. In particular, \cite{Maric_Dabora}, \cite{Dabora_Maric} propose an interference-forwarding strategy which is similar to the one used for the Type I channel in this paper.
In a similar setup, the works of Ng et al.
\cite{Jindal_cooperation} and H{\o}st-Madsen
\cite{Host-Madsen_cooperation} study the interference channel
with analog relay links at the receiver, and use the
compress-and-forward relay strategy to obtain capacity
bounds and asymptotic results.  %Compress-and-forward is also shown to
%be sum-capacity achieving in different asymptotic regimes for the Type
%I and Type II channels in this paper.

This paper is closely related to the work of Wang and Tse
\cite{Wang_ReceiverCooperation}, Prabhakaran and Viswanath
\cite{Promond_DestinationCooperation}, and Simeone et al.
\cite{simeone}, where the interference channel with limited receiver
cooperation is studied. In \cite{simeone}, the achievable rates of a
Wyner-type cellular model with either uni- or bidirectional
finite-capacity backhaul links are characterized. In
\cite{Wang_ReceiverCooperation}, a more general channel model in which
a two-user Gaussian interference channel is augmented with bidirectional
digital relay links is considered, and a conferencing protocol based
on the quantize-map-and-forward strategy of \cite{Avestimehr_relaynetwork}
is proposed.

The present paper considers a special case of the channel model in
\cite{Wang_ReceiverCooperation}, i.e.,
a simplified Gaussian Z-interference channel model with
a unidirectional digital relay link.  By
focusing on this special case, we are able
to derive concrete achievability results and upper bounds and obtain
insights on the rate improvement due to the relay link.
For example, while \cite{Wang_ReceiverCooperation} adopts a universal
power splitting ratio of \cite{Tse2007} at the transmitter to achieve
the capacity region to within 2 bits, this paper adapts the
power splitting ratio to channel parameters, and shows that
%In particular,
%this paper utilizes the classic decode-and-forward and
%compress-and-forward relay strategies and derives the {\it{exact}}
%capacity region in the strong interference regime, and gives an
%achievable rate region that is different from the one derived in
%\cite{Wang_ReceiverCooperation} in the weak interference regime.
%Another key difference between this paper and
%\cite{Wang_ReceiverCooperation} lies in the input strategy. Although
%Han-Kobayashi type of input is used in both papers, this paper,
%however, does not follow the Etkin-Tse-Wang \cite{Tse2007} power
%splitting strategy used in \cite{Wang_ReceiverCooperation}. This paper
%varies the power allocated to the private message and the common
%message and obtain an achievable rate region by time-sharing. This
%makes a remarkable different from \cite{Wang_ReceiverCooperation}
%where fixed common-private power ratio is adopted and no time-sharing
%is performed. In addition, with a unidirectional digital link, this
%paper shows that the direction of the link is crucial. For example,
in the weak interference regime a relay link from the interference-free
receiver to the interfered receiver is much more beneficial than a
relay link in the opposite direction for a Z-interference channel.

%When having bidirectional digital links, the relay strategies in this
%paper can be naturally extended for conferencing. For example, for
%one-round conferencing, receiver $2$ can partially decode-and-forward
%to receiver $1$ as in the type I channel, and receiver $1$ helps
%receiver $2$ using a combination of the decode-and-forward scheme and
%the compress-and-forward scheme as in the type II channel.
%Multiple-rounds conferencing is also possible. However, it is out of
%the scope of this paper.

\subsection{Outline of the Paper}

The rest of this paper is organized as follows. Section II presents
achievability results for the Type I Gaussian Z-relay-interference
channel using the decode-and-forward strategy.  Capacity
results are established for the strong interference regime; asymptotic sum-capacity result is established for the weak interference regime in
the high SNR/INR limit. Section III presents achievability results for
the Type II Gaussian Z-relay-interference channel using a combination
of the decode-and-forward scheme and the compress-and-forward scheme.
Capacity results are derived in the strong interference regimes;
asymptotic sum-capacity result is established for all channel
parameters in the limit of large relay link rate.  Section IV contains
concluding remarks.

\section{Gaussian Z-Interference Channel with \\
	a Relay Link: Type I}

\subsection{Channel Model and Notations}

The Gaussian Z-interference channel is modeled as follows
(see Fig.~\ref{Gaussian_Z_introduction}(a)):
\begin{equation} \label{input_output}
\left\{
  \begin{array}{l}
Y_1=h_{11}X_1 + h_{21}X_2 +Z_1 \\
Y_2=h_{22}X_2 + Z_2
  \end{array}
\right.
\end{equation}
where $X_1$ and $X_2$ are the transmit signals with power
constraints $P_1$ and $P_2$ respectively, $h_{ij}$ represents the
real-valued
channel gain from transmitter $i$ to receiver $j$, and $Z_1$, $Z_2$
are the independent additive white Gaussian noises (AWGN) with power $N$.
In addition, the Type I Gaussian Z-relay-interference channel is
equipped with a digital noiseless link of fixed capacity $R_0$ from
receiver 2 to receiver $1$.

Each transmitter $i$ independently encodes a message $m_i$ into a
codeword
$X_i^n(m_i)$ using a codebook $\mathcal{C}_{i}^n$ of $2^{nR_i}$
length-$n$ codewords satisfying an average power constraint $P_i$.
%The noiseless link between two receivers operate in a causal fashion.
Let $V^n$ be the output of the digital link from receiver $2$ to
receiver $1$ taken from a relay codebook $\mathcal{C}_R^n$, where
$|\mathcal{C}_R^n| \le 2^{nR_0}$. Receiver $1$ uses a decoding
function $\hat{m}_1 = f_{1}^n(Y_1^n, V^n)$. Receiver $2$ uses
a decoding function $\hat{m}_2 = f_{2}^n(Y_2^n)$.  The average
probability of error for user $i$ is defined as $P_{e,i}^n =
\mathbb{E}\left[\textrm{Pr}(\hat{m}_i \neq m_i)\right]$.
A rate pair $(R_1, R_2)$ is said to be achievable if for every $\epsilon
> 0$ and for all sufficiently large $n$, there exists a family of
codebooks $(\mathcal{C}_i^n, \mathcal{C}_R^n)$,
and decoding functions $f_i^n$, $i=1, 2$,
such that $\max_i \{P_{e,i}^n\} < \epsilon$. The capacity region is
defined as the set of all achievable rate pairs.

%\begin{figure} [t]
%\centering \centering \psfrag{X1}{$X_1$} \psfrag{X2}{$X_2$}
%\psfrag{Y1}{$Y_1$} \psfrag{Y2}{$Y_2$} \psfrag{Z1}{$Z_1$}
%\psfrag{Z2}{$Z_2$} \psfrag{h11}{$h_{11}$} \psfrag{h22}{$h_{22}$}
%\psfrag{h21}{$h_{21}$} \psfrag{R0}{$R_0$}
%\includegraphics[width=1.8in]{./figures/relay_ic_type_i}
%\caption{Gaussian Z-interference channel with a digital relay link}
%\label{Z_Relay}
%\end{figure}
%

To simplify the notation, the following definitions are used
throughout this paper:
\begin{eqnarray}
\mathsf{SNR_1} = \frac{|h_{11}|^2P_1}{N}  & &
\mathsf{SNR_2} = \frac{|h_{22}|^2P_2}{N}  \nonumber \\
\mathsf{INR_2} = \frac{|h_{21}|^2P_2}{N}  & &
\gamma (x)  =  \frac{1}{2} \log (1 + x) \nonumber
\end{eqnarray}
where $\log (\cdot)$ is base 2.
In addition, denote $\overline{\beta}=1-\beta$, and let
$(x)^{+} = \max \{x, 0\}$.

\subsection{Achievable Rate Region}

%Given $Y_2$'s observation, how should the relay link be used to help
%$Y_1$ decode $X_1$? Clearly, the decoding of $X_1$ is not possible, as
%$X_1$ is not observed at $Y_2$. Instead, $Y_2$ observes a noisy
%version of the interference at $Y_1$. Thus, compress-and-forward may
%be a sensible strategy in which $Y_2$ is described to $Y_1$ using a
%rate-constrained quantization codebook.  However, compress-and-forward
%does not take into account the fact that $X_2$ is not a Gaussian
%random noise but a codeword from a structured codebook, and that one
%may intentionally design $X_2$ in order to facilitate interference
%subtraction.

This paper uses a combination of the Han-Kobayashi common-private
power splitting technique and a decode-and-forward strategy for the
Gaussian Z-relay-interference channel, in which a common information
stream is decoded at receiver $2$, then binned and forwarded to
receiver $1$ for subtraction. %We
%call such a relay strategy partial decode-and-forward.
The main result of this section is the following achievability theorem.

\begin{theo} \label{theorem_z_type_I}
For the Type I Gaussian Z-interference channel with a digital relay
link of limited rate $R_0$ from the interference-free receiver to the
interfered receiver as shown in Fig.~\ref{Gaussian_Z_introduction}(a),
in the weak interference regime defined by $0 \le \mathsf{INR_2} <
\min\{\mathsf{SNR_2, INR_2^{*}}\}$, the following rate region is
achievable:
\begin{multline} \label{rate_region_Z_low}
\bigcup_{0 \leq \beta \leq 1}  \left \{ (R_1, R_2) \left| R_1 \leq \gamma
\left ( \frac{\mathsf{SNR}_1}{1 + \beta \mathsf{INR}_2}
\right), \right. \right. \\
R_2 \leq \min \left \{ \gamma (\mathsf{SNR}_2), \gamma(\beta
\mathsf{SNR}_2) +  \right. \\
\left. \left. \gamma \left(\frac{\overline{\beta}\mathsf{INR}_2}{1 +
\mathsf{SNR}_1 + \beta \mathsf{INR}_2} \right) +R_0 \right\}
   \right\},
\end{multline}
where
\begin{equation} \label{INR_2_star}
\mathsf{INR_2^*}= \left( (1+
\mathsf{SNR}_1)(2^{-2R_0}\mathsf{(1+SNR_2)} -1) \right)^+.
\end{equation}
In the strong interference regime defined by $\min\{\mathsf{
SNR_2, INR_2^{*}}\} \le \mathsf{INR_2} < \mathsf{INR}_2^{*}$,
the capacity region is given by
\begin{equation} \label{capacity_region_Z_moderate}
\left\{ (R_1, R_2) \left|
  \begin{array}{rll}
R_1 &\le& \gamma (\mathsf{SNR_1}) \\
R_2 &\le& \gamma (\mathsf{SNR_2})  \\
R_1+R_2 &\le& \gamma(\mathsf{SNR_1 + INR_2}) +R_0
  \end{array}
\right. \right \}.
\end{equation}
In the very strong interference regime defined by
$\mathsf{INR_2 \ge INR_2^{*} }$,
the capacity region is given by
\begin{equation} \label{capacity_region_Z_strong}
\left\{ (R_1, R_2) \left|
  \begin{array}{l}
R_1 \le \gamma (\mathsf{SNR_1}) \\
R_2 \le \gamma (\mathsf{SNR_2})  \\
  \end{array}
\right.  \right \}.
\end{equation}
\end{theo}

\begin{figure} [t]
\centering \psfrag{X1}{$X_1$} \psfrag{X2}{$X_2$} \psfrag{Y1}{$Y_1$}
\psfrag{Y2}{$Y_2$} \psfrag{U2}{$U_2$} \psfrag{W2}{$W_2$}
\psfrag{Z1}{$Z_1$} \psfrag{Z2}{$Z_2$} \psfrag{R0}{$R_0$}
\psfrag{S1}{$S_1$} \psfrag{T2}{$T_2$} \psfrag{S2}{$S_2$}
\psfrag{P1}{$P_1$} \psfrag{b_P2}{$\overline{\beta} P_2$}
\psfrag{bP2}{$\beta P_2$} \psfrag{h11}{$h_{11}$}
\psfrag{h22}{$h_{22}$} \psfrag{h21}{$h_{21}$}
\includegraphics[width=3.3in]{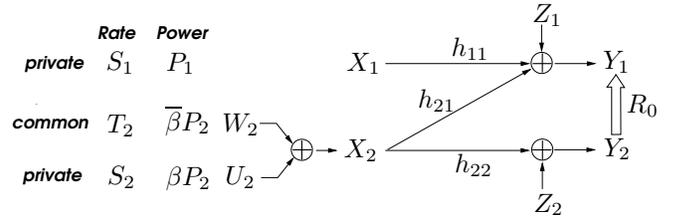}
\caption{Common-private power splitting for Type I channel.}
\label{power_splitting_Z}
\end{figure}

\begin{proof} %(\textbf{Achievability}):
We use the Han-Kobayashi \cite{HK1981} common-private power splitting scheme with Gaussian inputs to prove the achievability of
the rate regions (\ref{rate_region_Z_low}),
(\ref{capacity_region_Z_moderate}) and
(\ref{capacity_region_Z_strong}). As depicted in
Fig.~\ref{power_splitting_Z}, user $1$'s signal $X_1$ is intended for
decoding at $Y_1$ only.  User $2$'s signal $X_2$ is the superposition
of the private message $U_2$ and the common message $W_2$, i.e., $X_2 = U_2
+W_2$. The private message can only be decoded by the intended
receiver $Y_2$, while the common message can be decoded by both
receivers. Independent Gaussian codebooks of sizes $2^{nS_1}$,
$2^{nS_2}$ and $2^{nT_2}$ are generated according to i.i.d.\ Gaussian
distributions $X_1 \sim \mathcal{N}(0,P_1)$, $U_2 \sim \mathcal{N}
(0,\beta P_2)$, and $W_2 \sim \mathcal{N}(0,\overline{\beta}P_2)$,
respectively, where $0 \le \beta \le 1$. The encoded sequences $X_1^n$
and $X_2^n=U_2^n + W_2^n$ are then transmitted over a block of $n$
time instances.

Decoding takes place in two steps. First, $(W_2^n,U_2^n)$ are decoded at receiver $2$. The set of achievable rates $(T_2,S_2)$ is the capacity
region of a Gaussian multiple-access channel, denoted here by
$\mathcal{C}_2$, where
\begin{equation} \label{start}
\left\{
 \begin{array}{rll}
T_2 &\le& \gamma (\overline{\beta} \mathsf{SNR_2})  \\
S_2 &\le&  \gamma (\beta \mathsf{SNR_2})  \\
S_2+T_2  &\le&  \gamma (\mathsf{SNR_2}).
  \end{array}
\right.
\end{equation}
After $(W_2^n,U_2^n)$ are decoded at receiver $2$, $(X_1^n,W_2^n)$ are then decoded at receiver $1$ with $U_2^n$ treated as noise, but with the help of the relay link.
This is a multiple-access channel with a rate-limited relay $Y_2^n$, who
has complete knowledge of $W_2^n$. This channel is a special case of the
multiple-access relay channel studied in
\cite{Kramer_MARC_1} and \cite{Kramer_MARC_2}.
It is straightforward to show that a decode-and-forward relay strategy
is capacity achieving in this special case and its capacity region
$\mathcal{C}_1$ is the set of $(S_1,T_2)$ for which
\begin{equation} \label{end}
\left\{
 \begin{array}{rll}
S_{1} &\le&  \gamma \left(
    \D \frac{\mathsf{SNR_1}}{1 + \beta \mathsf{INR_2}}
    \right) \\
T_{2} &\le& \gamma \left(
    \D \frac{\overline{\beta} \mathsf{INR_2}}
    {1+ \beta \mathsf{INR_2}} \right) +R_0 \\
S_{1}+T_{2} &\le &\gamma \left(
    \D \frac{\mathsf{SNR_1} + \overline{\beta}
    \mathsf{INR_2}}{1 + \beta \mathsf{INR_2}} \right) +R_0.
  \end{array}
\right.
\end{equation}

%Since all $S_1$, $S_{2}$, and $T_{2}$ appear in (\ref{start}) and
%(\ref{end}) with coefficient zero or one, at the maximal point
%decreasing $T_{2}$ by a small $r>0$ increase each of the others by
%exact $r$ or zero. Therefore, by noting that $R_2=S_{2}+T_{2}$, we
%conclude that $\mathfrak{R}$ is pentagon which is delimited by
%straight lines of slope $0, -1, \infty$. Therefore,

%\begin{figure} [t]
%\centering \psfrag{S1}{$S_1$} \psfrag{S2}{$S_2$} \psfrag{T2}{$T_2$}
%\psfrag{O}{$0$} \psfrag{delta}{$\delta$}
%\psfrag{C1}{$\mathcal{C}_1$} \psfrag{C2}{$\mathcal{C}_2$}
%\includegraphics[width=2.2in]{./figures/3dim}
%\caption{Incorporating $\mathcal{C}_1$ and $\mathcal{C}_2$ for Type I
%channel.}
%\label{3dim}
%\end{figure}
%
%\begin{figure} [t]
%\centering \psfrag{S1}{$S_1$} \psfrag{S2+T2}{$S_2+T_2$}
%\psfrag{O}{$0$} \psfrag{delta}{$\delta$}
%\includegraphics[width=1.8in]{./figures/r1r2plain}
%\caption{Pentagon achievable rate region.} \label{R1R2plain}
%\end{figure}

An achievable rate region of the Gaussian Z-interference channel with
a relay link is then the set of all $(R_1,R_2)$ such that $R_1=S_1$
and $R_2=S_{2}+T_{2}$ for some $(S_1, T_2) \in \mathcal{C}_1$ and
$(S_2, T_2) \in \mathcal{C}_2$. Further, since $\mathcal{C}_1$ and
$\mathcal{C}_2$ depend on the common-private power splitting ratio
$\beta$, the convex hull of the union of all such $(R_1,R_2)$ sets
over all choices of $\beta$ is achievable.

%In the following, we first show that for each fixed $\beta$, the set
%of achievable rates $(R_1,R_2)$ is a pentagon. We then show that the
%convex hull of the union of these pentagons reduces to the regions
%(\ref{rate_region_Z_low}), (\ref{capacity_region_Z_moderate}) and
%(\ref{capacity_region_Z_strong}).
%
%Fig.~\ref{3dim} illustrates an example of a $(S_2,T_2)$-pentagon
%(\ref{start}) and a $(S_1,T_2)$-pentagon (\ref{end}) in a
%three dimensional diagram.  As $S_1$ moves from its maximal value
%downward (i.e.\ from $A$ to $C$), the maximal achievable $T_2+S_2$
%increases from the sum rate corresponding to $A'$ to that
%corresponding to $C'$.  The sum
%rate $T_2+S_2$ is a constant beyond $C'$, because the line segment
%$C'D'$ is at 45 degrees.  Now, since the line segment $AC$ is also
%at 45 degrees, from $A$ to $C$, the sum $S_1+T_2$ is a constant as
%well. Consequently $S_1+T_2+S_2$ is also a constant from $A$ to $C$.
%Therefore, when plotting $R_2$ (which is $S_2+T_2$) vs.\ $R_1$
%(which is $S_1$), we obtain a pentagon shape with a 45-degree edge
%$A''C''$ as shown in Fig.~\ref{R1R2plain}. Beyond point $C$,
%$S_2+T_2$ stays as a constant as $S_1$ decreases.

A Fourier-Motzkin elimination method
(see e.g.\ \cite{Chong2006}) can be used to show that for each fixed $\beta$,
the achievable $(R_1,R_2)$'s form a pentagon region characterized by
\begin{equation} \label{pentagon_region}
%\bigcup_{0 \le \beta \le 1}
\mathcal{R}_{\beta}=
\left\{ (R_1, R_2) \left|
 \begin{array}{l}
\D
R_1 \le \gamma \left(\mathsf{\frac{SNR_1}{1 + \beta INR_2}} \right) \\
\D
R_2 \le \min \{\gamma(\mathsf{SNR_2}), \gamma(\mathsf{\beta SNR_2}) + \\
\D
\qquad \qquad \quad  \gamma \left(\mathsf{\frac{\overline{\beta} INR_2}
    {1 +\beta INR_2}} \right) + R_0  \} \\
\D
R_1+R_2 \le \gamma(\mathsf{\beta SNR_2}) +  \\
\qquad \qquad \quad \gamma \D \left( \mathsf{\frac{SNR_1 +
\overline{\beta}INR_2}{1 + \beta INR_2}} \right) +R_0
\\
\end{array}
\right. \right \}.
\end{equation}
The convex hull of the union of these pentagons over $\beta$ gives the
complete achievability region. It happens that the union of the pentagons,
i.e.\ $\bigcup_{0 \le \beta \le 1} \mathcal{R}_{\beta}$, is already convex.
Therefore, convex hull is not needed. In the following, we give an explicit
expression for $\bigcup_{0 \le \beta \le 1} \mathcal{R}_{\beta}$.
%\begin{equation}
%\bigcup_{0 \le \beta \le 1} \mathcal{R}_{\beta} =
%\mathrm{co} \left\{ \bigcup_{0 \le \beta \le 1} \mathcal{R}_{\beta} \right\}
%\end{equation}
%where ``$\mathrm{co}$'' denotes convex hull. This allows
%and give an explicit expression for the resulting achievable region as a function of the
%channel parameters.

\begin{figure} [t]
\centering \psfrag{beta1}{$\beta=1$} \psfrag{beta0}{$\beta=0$}
\psfrag{betastar}{$\beta^*$} \psfrag{R1}{$R_1$}  \psfrag{R2}{$R_2$}
\psfrag{0}{$0$} \psfrag{A}{$\gamma(\mathsf{\frac{SNR_1}{1+INR_2}})$}
\psfrag{C}{$\mathsf{\gamma(SNR_1)}$}
\psfrag{D}{$\mathsf{\gamma(INR_2)} + R_0$}
\psfrag{B}{$\mathsf{\gamma(SNR_2)} +R_0$}
\psfrag{E}{$\mathsf{\gamma(SNR_2)}$}
\psfrag{F}{$\gamma \left( \mathsf{\frac{INR_2}{1+SNR_1}} \right) + R_0$}
\includegraphics[width=3.0in]{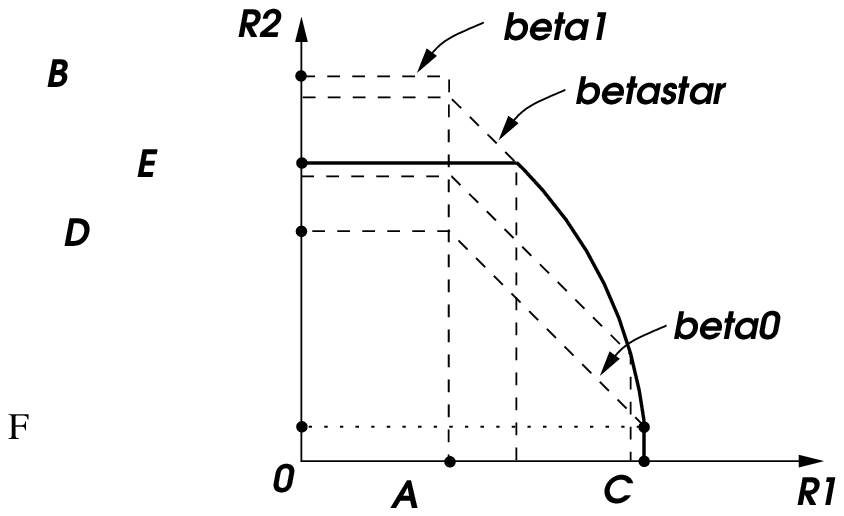}
\caption{The union of rate region pentagons when $\mathsf{INR_2 \le SNR_2}$.} \label{pentagon_union_weak}
\end{figure}

Consider first the regime where $\mathsf{INR_2}
\le \mathsf{SNR_2}$. Ignore for now the constraint $R_2 \le
\gamma(\mathsf{SNR_2})$ and focus on an expanded pentagon defined by
$\{(R_1,R_2) \left| R_1 \le f_1(\beta), R_2 \le f_2(\beta),
R_1+R_2 \le f_3(\beta) \right.\}$, where $f_1(\beta)$ is the
$R_1$ constraint in (\ref{pentagon_region}), $f_2(\beta)$
is the second term of the min expression in the $R_2$ constraint in
(\ref{pentagon_region}),
and $f_3(\beta)$ is the $R_1+R_2$ constraint in (\ref{pentagon_region}).
%\begin{eqnarray}
%f_1(\beta) &=& \gamma \left(\mathsf{\frac{SNR_1}
%    {1 + \beta INR_2}} \right) \\
%f_2(\beta) &=& \gamma \left(\mathsf{\beta SNR_2}\right) +
%    \gamma \left(\mathsf{\frac{\overline{\beta} INR_2}
%        {1 +\beta INR_2}} \right) + R_0 \\
%f_3(\beta) &=& \gamma \left(\mathsf{\beta SNR_2}\right) +
%    \gamma \left( \mathsf{\frac{SNR_1 +
%    \overline{\beta}INR_2}{1 + \beta INR_2}} \right) +R_0
%\end{eqnarray}

It is easy to verify that when $\beta=1$, the expanded pentagon
reduces to a rectangular region, as shown in
Fig.~\ref{pentagon_union_weak}. Further, as $\beta$ decreases from 1
to 0, $f_1(\beta)$ monotonically increases and both $f_2(\beta)$
and $f_3(\beta)$ monotonically decrease, while
$f_2(\beta)-f_3(\beta)$ remains a constant in the regime where
$\mathsf{INR_2 \le SNR_2}$. Since $f_1(\beta)$, $f_2(\beta)$ and
$f_3(\beta)$ are all continuous functions of $\beta$, as $\beta$
decreases from $1$ to $0$, the upper-right corner point of the
expanded pentagon moves vertically downward in the $R_2-R_1$ plane,
while the lower-right corner point moves downward and to the right
in a continuous fashion.  Consequently, the union of these expanded
pentagons is defined by $R_1 \le \gamma(\mathsf{SNR_1})$, $R_2 \le
\gamma(\mathsf{SNR_2})+R_0$, and lower-right corner points of the
pentagons $(R_1,R_2)$ with
\begin{equation}
\left\{
\begin{array}{lll}
R_1 & = & \gamma \left ( \D \frac{\mathsf{SNR}_1}{1 +
    \beta \mathsf{INR}_2} \right) \\
R_2 & = & \gamma(\beta \mathsf{SNR}_2) +
    \gamma \left(\D \frac{\overline{\beta}\mathsf{INR}_2}{1 +
    \mathsf{SNR}_1 + \beta \mathsf{INR}_2} \right) +R_0
\end{array}
\right.
\label{convex_region_chapter_2}
\end{equation}
where $0\le \beta \le 1$.
We prove in Appendix \ref{appendix_convexity} that such a region
is convex when $\mathsf{INR_2 \le SNR_2}$. Thus, convex hull is not needed.
Finally, incorporating the constraint $R_2 \le \gamma(\mathsf{SNR_2})$
gives the achievable region (\ref{rate_region_Z_low}).

\begin{figure} [t]
\centering \psfrag{beta1}{$\beta=1$} \psfrag{beta0}{$\beta=0$}
\psfrag{betastar}{$\beta^*$} \psfrag{R1}{$R_1$}  \psfrag{R2}{$R_2$}
\psfrag{0}{$0$} \psfrag{A}{$\gamma(\mathsf{\frac{SNR_1}{1+INR_2}})$}
\psfrag{C}{$\mathsf{\gamma(SNR_1)}$}
\psfrag{D}{$\mathsf{\gamma(INR_2)}+R_0$}
\psfrag{B}{$\mathsf{\gamma(SNR_2)} +R_0$}
\psfrag{E}{$\mathsf{\gamma(SNR_2)}$}
\psfrag{F}{$\gamma \left( \mathsf{\frac{INR_2}{1+SNR_1}} \right) + R_0$}
\includegraphics[width=2.7in]{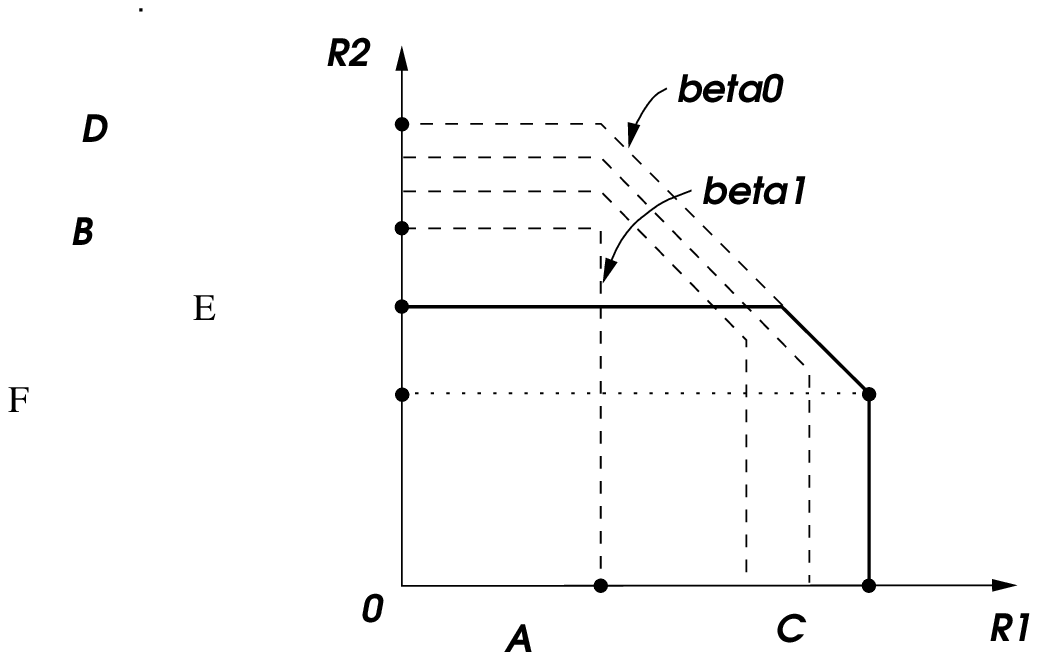}
\caption{The union of rate region pentagons when $\mathsf{INR_2 \ge SNR_2}$.} \label{pentagon_union_strong}
\end{figure}

Now, consider the regime where $\mathsf{INR_2
\ge SNR_2}$. In this regime, $f_1(\beta)$, $f_2(\beta)$ and
$f_3(\beta)$ are all increasing functions as $\beta$ goes from 1 to
0. Consequently, $\bigcup_{0 \le \beta \le 1} \mathcal{R}_{\beta} =
\mathcal{R}_{0}$, as illustrated in
Fig.~\ref{pentagon_union_strong}. Therefore, convex hull is not needed.
Thus, the achievable rate region
simplifies to
\begin{equation}
\left\{ (R_1, R_2) \left|
  \begin{array}{l}
R_1 \le \gamma (\mathsf{SNR_1}) \\
R_2 \le \min \{ \gamma (\mathsf{SNR_2}), \mathsf{\gamma(INR_2)} +R_0 \}\\
R_1+R_2 \le \gamma(\mathsf{SNR_1 + INR_2}) +R_0
  \end{array}
\right.  \right \},
\end{equation}
which is equivalent to (\ref{capacity_region_Z_moderate}) by noting
that
\begin{equation} \label{INRstarcondition}
\mathsf{\gamma(INR_2)} +R_0 \ge \gamma (\mathsf{SNR_2})
\end{equation}
when $\mathsf{INR_2 \ge  SNR_2}$.

We have so far obtained the achievable rate regions for the regimes
$\mathsf{INR_2 \le SNR_2}$ and $\mathsf{INR_2 \ge SNR_2}$ as in
(\ref{rate_region_Z_low}) and (\ref{capacity_region_Z_moderate})
respectively. Both expressions can be further simplified in some
specific cases. Inspecting Figs.~\ref{pentagon_union_weak} and
\ref{pentagon_union_strong}, it is easy to see that when
$\mathsf{INR_2 \ge INR_2^*}$, where $\mathsf{INR_2^*}$ is as
defined in (\ref{INR_2_star}), the horizontal line $R_2 =
\gamma(\mathsf{SNR_2})$ is below the lower-right corner point
corresponding to $\beta =0$, i.e.,
\begin{equation}
\gamma(\mathsf{SNR_2}) \le  \gamma \D \left(
\mathsf{\frac{INR_2}{1+SNR_1}} \right) + R_0.
\end{equation}
Therefore, in both the $\mathsf{INR_2 \le SNR_2}$
(Fig.~\ref{pentagon_union_weak}) and the $\mathsf{INR_2 \ge SNR_2}$
(Fig.~\ref{pentagon_union_strong}) regimes, whenever $\mathsf{INR_2} \ge
\mathsf{INR_2^*}$, the achievable rate region reduces to a rectangle
as in (\ref{capacity_region_Z_strong}). This is the very strong
interference regime.

Noting the fact that  $\mathsf{INR_2^*}$ can be greater or less than
$\mathsf{SNR_2}$ depending on $R_0$, we see that the achievability
result for the Type I channel is divided into the weak, strong, and
very strong interference regimes as in
(\ref{rate_region_Z_low}), (\ref{capacity_region_Z_moderate}) and
(\ref{capacity_region_Z_strong}) respectively.

Finally, it is possible to prove a converse in the strong and very
strong interference regimes. The converse proof is presented in
Appendix \ref{appendix_strong_I}.
\end{proof}

It is important to note that the achievable region of Theorem
\ref{theorem_z_type_I} is derived assuming fixed powers $P_1$ and
$P_2$ at the transmitters.  It is possible that time-sharing among
different transmit powers may enlarge the achievable rate
region. For simplicity in the presentation of closed-form expressions
for achievable rates, time-sharing is not explicitly incorporated in
the achievability theorems in this paper.

\subsection{Numerical Examples}

%Note that, time-sharing technique is conceivable of potentially enlarging
%the achievable rate region. However, we will not incorporate time-sharing
%into the achievable scheme in the following theorems, because it is very hard to
%characterize time-sharing into closed form expressions.

%The achievability results for the classic Gaussian interference
%channel are categorized to the weak interference, the strong
%interference, and the very strong interference regimes. The capacity
%region of the classic interference channel in the weak interference
%case is still open. In the strong interference case, the capacity
%region is known as a pentagon. In the very strong interference case,
%the capacity region becomes a rectangle. The result of the previous
%section shows that the capacity region of the Type I Gaussian
%Z-relay-interference channel follows the same pattern.

%It is interesting to note
%that the relay link does not change the boundary between the weak and
%the strong interference regimes, but it does change the boundary
%between the strong and the very strong interference regimes. In other
%words, receiver-side relaying may potentially turn a strong
%interference channel into a very strong interference channel, but it
%never turns a weak interference channel into a strong interference
%channel.

It is instructive to numerically compare the achievable regions of
the Gaussian Z-interference channel with and without the relay link.
First, observe that when $R_0=0$, the achievable rate region
(\ref{rate_region_Z_low}) and the capacity region results
(\ref{capacity_region_Z_moderate}) (\ref{capacity_region_Z_strong})
reduce to previous results obtained in \cite{HK1981} and \cite{Sato}.

\begin{figure} [t]
\centering
\includegraphics[width=3.5in]{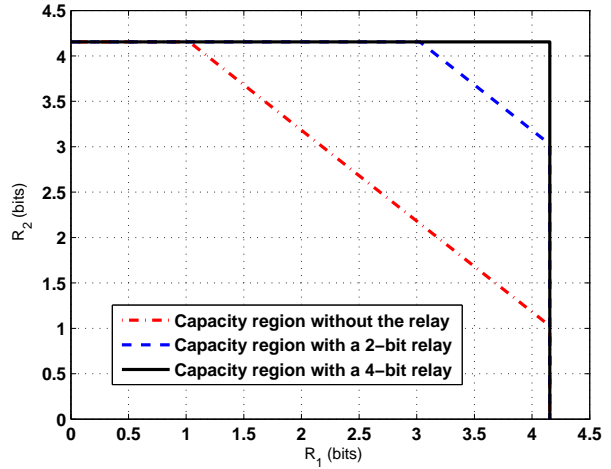}
\caption{Capacity region of the Gaussian Z-interference channel in
the strong interference regime with and without a digital relay link
of Type I.}
\label{pic_capacity_region_Z}
\end{figure}

In the strong and very strong interference regimes, the capacity
region of a Type I Gaussian Z-relay-interference channel is achieved by transmitting common information only at $X_2$. In
the very strong interference regime, the relay link does not increase
capacity, because the interference is already completely decoded and
subtracted, even without the help of the relay. In the strong
interference regime, the relay link increases the capacity by helping
the common information decoding at $Y_1$. In fact, a relay link of
rate $R_0$ increases the sum capacity by exactly $R_0$ bits.
As a numerical example, Fig.~\ref{pic_capacity_region_Z} shows the
capacity region of a Gaussian Z-interference channel in the strong
interference regime with and without the relay link. The channel
parameters are set to be $\mathsf{SNR}_1=\mathsf{SNR}_2=25$dB,
$\mathsf{INR}_2=30$dB. The capacity region without the relay is the
dash-dotted pentagon. With $R_0=2$ bits, the capacity region expands to the
dashed pentagon region, which represents an increase in sum rate of
exactly 2 bits.  As $R_0$ increases to $4$ bits, the channel falls
into the very strong interference regime. The capacity region becomes
the solid rectangular region.

%{\em \textbf{Further, the increase in sum capacity can be arbitrarily
%divided between the two users, as long as the individual rates are
%below their respectively interference-free upper bound.  This is
%because the relay link can either increase the common information rate
%(which improve the total rate at the interference-free receiver), or
%increase the power of the common information component of $X_2$
%(which decreases the noise at the interfered receiver), or do a
%combination of both. }}

\begin{figure} [t]
\centering
\includegraphics[width=3.5in]{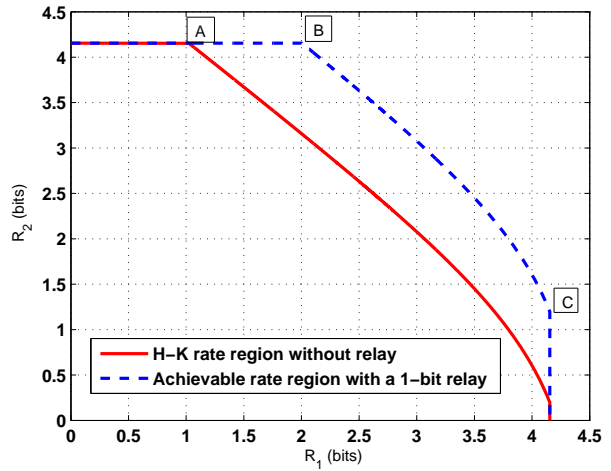}
\caption{Achievable rate region of the Gaussian Z-interference channel
in the weak interference regime with and without a digital relay link
of Type I.}
\label{pic_rate_region_Z}
\end{figure}

In the weak interference regime, the achievable rate region in Theorem
\ref{theorem_z_type_I} is obtained by a Han-Kobayashi
common-private power splitting scheme.  By inspection, the effect of a
relay link is to shift the rate region curve upward by $R_0$ bits
while limiting $R_2$ by its single-user bound $\gamma(\mathsf{SNR}_2)$.
Interestingly, although the relay link of rate $R_0$ is provided from receiver $2$ to receiver $1$, it can help $R_2$ by exactly
$R_0$ bits, while it can only help $R_1$ by strictly less than $R_0$ bits.
%
%{\em \textbf{The reason for this phenomenon is as follows. For the
%relay to help increase $R_2$ while keeping $R_1$ fixed, $X_2$ must
%increase the common information rate. Since the common information is
%known at $Y_2$, relaying from $Y_2$ to $Y_1$ by binning achieves the
%cut-set bound---every bit in the relay link is worth one bit in $R_2$.
%On the other hand, for the relay to help increase $R_1$ while keeping
%$R_2$ fixed, $X_2$ needs to convert some of the private information into
%common information. In other words, the relay needs to use $R_0$ to
%describe a larger portion of $X_2$, (which reduces the interference
%for user 1).  The benefit of interference reduction to $R_1$ is in
%general strictly less than $R_0$. But, as the next section shows,
%the benefit is asymptotically equal to $R_0$ in the high SNR and
%high INR limit.}}
%
As a numerical example, Fig.~\ref{pic_rate_region_Z} shows the
achievable rate region of a Gaussian Z-interference channel
with $\mathsf{SNR}_1=\mathsf{SNR}_2=25$dB
and $\mathsf{INR}_2=20$dB. The solid curve represents the rate region
achieved without the relay link. The dashed rate region is with a relay of rate $R_0=1$ bit. For most part of the curve, $R_0$
provides a 1-bit increase in $R_2$, but a less than 1-bit increase in
$R_1$.

It is illustrative to identify the correspondence between the
various points in the rate region and the different common-private
splittings in the weak interference regime. Point $A$ corresponds to
$\beta=1$. This is where the entire $X_2$ is private message. In this
case, it is easy to verify that the first term of $R_2$ in
(\ref{rate_region_Z_low}) is less than the second term:
\begin{equation} \label{1st_less_than_2nd}
\gamma (\mathsf{SNR}_2) < \gamma(\beta \mathsf{SNR}_2) + \gamma
\left(\frac{\overline{\beta}\mathsf{INR}_2}{1 + \mathsf{SNR}_1 +
\beta \mathsf{INR}_2} \right)+ R_0
\end{equation}
As $\beta$ decreases, more private message is converted into common
message, which means that less interference is seen at receiver $1$.
As a result, $R_1$ increases, $R_2$ is kept at a constant (since
(\ref{1st_less_than_2nd}) continues to hold). Graphically, as $\beta$
decreases from $1$, the achievable rate pair moves horizontally from
point $A$ to the right until it reaches point $B$, corresponding to
some $\beta^*$, after which the second term of $R_2$ in
(\ref{rate_region_Z_low}) becomes less than the first term
$\gamma(\mathsf{SNR}_2)$. The value of $\beta^*$ can be computed as
\begin{equation} \label{beta_star}
\beta^* = \frac{(1+\mathsf{SNR}_1)(1+\mathsf{SNR}_2) - 2^{2R_0}(1 +
\mathsf{SNR}_1 + \mathsf{INR}_2)}{2^{2R_0}\mathsf{SNR}_2(1 +
\mathsf{SNR}_1 + \mathsf{INR}_2) - \mathsf{INR}_2(1 +
\mathsf{SNR}_2)}.
\end{equation}
As $\beta$ decreases further from $\beta^*$, more private message is
converted into common message, which makes $R_1$ even larger.
However, when $\beta < \beta^*$, the amount of common message can be
transmitted is restricted by the interference link $h_{21}$ and the
digital link rather than the direct link $h_{22}$.
Therefore, user $2$'s data rate cannot be kept as a constant;
$R_2$ goes down as user $1$'s rate goes up. As shown in
Fig.~\ref{pic_rate_region_Z}, the achievable rate pair moves from
point $B$ to point $C$ as $\beta$ decreases from $\beta^*$ to $0$.
Point $C$ corresponds to where the entire $X_2$ is
common message.

\subsection{Asymptotic Sum Capacity}

Practical communication systems often operate in the
interference-limited regime, where both the signal and the
interference are much stronger than noise.  In this section, we
investigate the asymptotic sum capacity of the Type I Gaussian
Z-relay-interference channel in the weak interference regime
%\begin{equation} \label{extreme_condition}
%\min \{\mathsf{SNR}_1, \mathsf{SNR}_2, \mathsf{INR_2}\} \gg 1.
%\end{equation}
%More precisely, we let noise power $N \rightarrow 0$, while keeping
where noise power $N \rightarrow 0$, while
power constraints $P_1$, $P_2$, channel gains $h_{ij}$, and
the digital relay link rate $R_0$ are kept fixed. In other words,
$\mathsf{SNR}_1, \mathsf{SNR}_2, \mathsf{INR_2} \rightarrow \infty$,
while their ratios are kept constant.

%\subsubsection{Asymptotic Capacity via Partial Interference
%Forwarding}

Denote the sum capacity of a Type I Gaussian Z-interference channel
with a relay link of rate $R_0$ by $C_{sum}(R_0)$.  Without the digital
relay link, or equivalently $R_0=0$, the sum capacity of the classic
Gaussian Z-interference channel in the weak interference regime
(i.e.\ $\mathsf{INR_2 \le SNR_2}$) is given
by \cite{Sason2004, Tse2007}:
\begin{equation} \label{define_C_sum0}
C_{sum}(0) =\gamma (\mathsf{SNR}_2) + \gamma \left(
\frac{\mathsf{SNR}_1}{1 + \mathsf{INR}_2} \right),
% \stackrel{\bigtriangleup}{=} C_{sum}(0)
\end{equation}
which is achieved by independent Gaussian codebooks and treating
the interference as noise at the receiver. In the high SNR/INR limit,
the above sum capacity becomes
\begin{equation} \label{Z-channel-sum_capacity}
C_{sum}(0) \approx \frac{1}{2} \log \left(
\frac{\mathsf{SNR_2}(\mathsf{SNR_1 +INR_2})}{\mathsf{INR}_2} \right),
\end{equation}
where the notation $f(x) \approx g(x)$ is used to denote $\lim f(x)
-g(x) = 0$. In the above expression, the limit is taken as $N
\rightarrow 0$.

Intuitively, with a digital relay link of finite capacity $R_0$,
the sum-rate increase due to the relay must be bounded by $R_0$. The
following theorem shows that in the high SNR/INR limit, the asymptotic sum-capacity increase is in fact $R_0$ in the weak-interference regime.

\begin{theo} \label{theorem_sum_capacity_typeI}
For the Type I Gaussian Z-interference channel with a digital relay
link of limited rate $R_0$ from the interference-free receiver to the
interfered receiver as shown in Fig.~\ref{Gaussian_Z_introduction}(a),
when $\mathsf{INR_2} \le \mathsf{\min \{SNR_2, INR_2^*\}}$, the
asymptotic sum capacity is given by
\begin{equation} \label{sum_capacity_typeI}
C_{sum}(R_0) \approx C_{sum}(0) + R_0.
\end{equation}
%where $C_{sum}(0)$ is defined in (\ref{define_C_sum0}).
\end{theo}

\begin{proof}
We first prove the achievability. As illustrated in
Fig.~\ref{pentagon_union_weak}
the sum rate of the Type I Gaussian Z-relay-interference channel
is achieved with $\beta=\beta^*$, where $\beta^*$ is as derived
in (\ref{beta_star}). In the high SNR/INR limit, we have
\begin{equation}
\lim_{N \rightarrow 0}\beta^* = \frac{2^{-2R_0}}{1 +
(1-2^{-2R_0})\frac{\mathsf{INR_2}}{\mathsf{SNR_1}} }.
\end{equation}
Substituting this $\beta^*$ into the achievable rate pair in
(\ref{rate_region_Z_low}), we obtain the asymptotic rate pair as
\begin{equation}
\left\{
  \begin{array}{l}
\D R_1 \approx \frac{1}{2} \log \left( 1 +
    \frac{\mathsf{SNR_1}}{\mathsf{INR_2}} \right) + R_0 \\
\D R_2 \approx \frac{1}{2} \log(\mathsf{SNR_2}) \end{array}
\right.
\end{equation}
which gives the following asymptotic sum rate:
\begin{eqnarray} \label{Z-relaychannel_sum_rate}
R_{sum} & \approx &\frac{1}{2} \log \left(
    \frac{\mathsf{SNR_2}(\mathsf{SNR_1 +INR_2})}
    {\mathsf{INR}_2} \right) + R_0 \nonumber \\
& \approx & C_{sum}(0) +R_0.
\end{eqnarray}

The converse proof starts with Fano's inequality. Denote the output
of the digital relay link over the $n$-block by $V^n$. Since the
digital link has a capacity limit $R_0$, $V^n$ is a discrete random
variable with $H(V^n) \le nR_0$. For a codebook of block length $n$,
we have
\begin{eqnarray}
\lefteqn{n(R_1 +R_2)} \nonumber \\
%&\stackrel{(a)}{\le} &
&\le &
I(X_1^n; Y_1^n, V^n) + I(X_2^n; Y_2^n) +
n\epsilon_n \nonumber \\
&=&I(X_1^n;Y_1^n) + I(X_1^n; V^n |Y_1^n)  + I(X_2^n; Y_2^n) +
n\epsilon_n \nonumber \\
%&\stackrel{(b)}{\le}
&{\le} &I(X_1^n;Y_1^n) + H(V^n |Y_1^n)  + I(X_2^n; Y_2^n) +
n\epsilon_n \nonumber \\
%&\stackrel{(c)}{\leq}
&{\leq}
&I(X_1^n;Y_1^n)+ I(X_2^n; Y_2^n) +  nR_0 +
n\epsilon_n \nonumber \\
&{\leq}&
%&\stackrel{(d)}{\leq}&
nC_{sum}(0) + nR_0 + n\epsilon_n,
\end{eqnarray}
%where
%\begin{description}
%  \item[(a)] follows from Fano's inequality;
%  \item[(b)] is due to $H(V^n|X_1^n,Y_1^n) \ge 0$;
%  \item[(c)] is due to $H(V^n|Y_1^n) \le H(V^n) \le n R_0$;
%  \item[(d)] follows from the definition of sum capacity.
%\end{description}
%
%Then, we have the following upper bound for the sum rate:
%\begin{equation}
%R_1+R_2 \le C_{sum}(0) +R_0 +  \epsilon_n
%\end{equation}
where $\epsilon_n \rightarrow 0$ as $n$ goes to infinity.  Note that
this upper bound holds for all ranges of $\mathsf{SNR_1}$,
$\mathsf{SNR_2}$, and $\mathsf{INR_2}$. This, when combined with the
asymptotic achievability result proved earlier, gives the asymptotic sum capacity $C_{sum}(R_0) \approx C_{sum}(0) +R_0$.
\end{proof}

%The above result shows that using the partial decode-and-forward
%strategy, every bit in the digital relay link is worth one bit to
%the sum capacity asymptotically in the high SNR and INR limit.

The above proof focuses on the sum-capacity achieving power splitting
ratio $\beta^*$. As $\beta \le \beta^*$, the achievable rate pair goes
from point $B$ to point $C$ along the dashed curve as shown in
Fig.~\ref{pic_rate_region_Z}. It turns out that for any fixed $0 <
\beta \le \beta^*$, the sum rate also asymptotically approaches the
upper bound, thus providing an alternative proof for Theorem
\ref{theorem_sum_capacity_typeI}.

To see this, fix some arbitrary $0 < \beta \le \beta^*$,
the sum rate corresponding to this $\beta$ is given in Theorem
\ref{theorem_z_type_I} as
\begin{eqnarray}
R_{sum} &=& \gamma \left ( \frac{\mathsf{SNR}_1}{1 + \beta
\mathsf{INR}_2} \right) + \gamma(\beta \mathsf{SNR}_2) +
\nonumber \\
&&  \qquad \gamma \left(\frac{\overline{\beta}\mathsf{INR}_2}{1 +
\mathsf{SNR}_1 + \beta
\mathsf{INR}_2} \right) +R_0 \nonumber \\
&=& \frac{1}{2} \log \left (\mathsf{\frac{1 + \beta SNR_2}{1 + \beta
INR_2}} \right) +   \gamma (\mathsf{SNR_1 + INR_2})  + R_0 \nonumber \\
&\approx& \frac{1}{2} \log \left(
\frac{\mathsf{SNR_2}(\mathsf{SNR_1 +INR_2})}{\mathsf{INR}_2} \right)
+ R_0,
\end{eqnarray}
which is the asymptotic sum capacity. This calculation implies that in
the high SNR/INR regime, the dashed curve in
Fig.~\ref{pic_rate_region_Z} has an initial slope of -1 as $\beta$
goes from $\beta^*$ to $0$.

%\subsubsection{Asymptotic Capacity via compress-and-forward}

Interestingly, decode-and-forward is not the only way to
asymptotically achieve the sum capacity of the Type I channel.
The following shows that a compress-and-forward relaying scheme,
although strictly suboptimal in finite SNR/INR, becomes asymptotically
sum-capacity achieving in the high SNR/INR limit in the weak
interference regime, thus giving yet another proof of Theorem
\ref{theorem_sum_capacity_typeI}.

%\begin{figure} [t]
%\centering \psfrag{X1}{$X_1$} \psfrag{X2}{$X_2$} \psfrag{Y1}{$Y_1$}
%\psfrag{Y2}{$Y_2$} \psfrag{Z1}{$Z_1$} \psfrag{Z2}{$Z_2$}
%\psfrag{h11}{$h_{11}$} \psfrag{h22}{$h_{22}$} \psfrag{h21}{$h_{21}$}
%\psfrag{Y2hat}{$\hat{Y}_2$} \psfrag{R0}{$R_0$}
%\includegraphics[width=2.0in]{./figures/quantize_relay_ic_type_i}
%\caption{compress-and-forward in Gaussian Z-interference relay
%channel.} \label{quantization}
%\end{figure}

In the compress-and-forward scheme, no common-private power splitting is performed. Each receiver only decodes the message
intended for it. Specifically, receiver
$2$ compresses its received signal $Y_2$ into $\hat{Y}_2$, then forwards it to receiver $1$ through the digital link $R_0$.

Clearly, the rate of user $2$ is given by
\begin{equation}  \label{R_2_quantization}
R_2 = \max_{p(x_2)}I(X_2; Y_2).
\end{equation}
Using the Wyner-Ziv coding strategy
\cite{WynerZiv,Cover1979}, for a fixed $p(x_2)$, the following
rate for user $1$ is achievable:
\begin{equation}\label{R_1_quantization}
R_1 = \max_{p(x_1)p(\hat{y}_2|y_2)}I(X_1; Y_1,\hat{Y}_2)
\end{equation}
under the constraint
\begin{equation} \label{Wyner-Ziv_constraint}
I(Y_2;\hat{Y}_2 | Y_1) \le R_0.
\end{equation}
The optimization in (\ref{R_1_quantization}) is in general hard.
%not computable without specifying
%he input distributions $p(x_1)$ and $p(x_2)$ and the quantization
%cheme $p(\hat{y}_2|y_2)$,
Here, we adopt independent Gaussian codebooks with $X_1
\sim \mathcal{N}(0,P_1)$ and $X_2 \sim \mathcal{N}(0,P_2)$, and a
Gaussian quantization scheme for the compression of $Y_2$:
\begin{equation} \label{Gaussian_noise_quantization}
\hat{Y}_2 =Y_2 + e
\end{equation}
where $e$ is a Gaussian random variable independent of $Y_2$, with a
distribution $ \mathcal{N}(0, \sigma^2)$. We show in Appendix
\ref{appendix_wyner_ziv} that this choice of
$p(x_1)p(x_2)p(\hat{y}_2|y_2)$ gives the following achievable rate
pair:
\begin{equation} \label{rate_pair_quantization}
\left\{
  \begin{array}{l}
\D R_1 = \gamma \left( \frac{\mathsf{SNR_1}}{1 + \mathsf{INR_2}}
\right) + R_0 - \delta_0(R_0) \\
\D R_2 = \gamma(\mathsf{SNR_2})
  \end{array}
\right.
\end{equation}
where
\begin{eqnarray}
\lefteqn{\delta_0(R_0) =} \nonumber \\
&&\gamma \left( \frac{(2^{2R_0}-1)(1+ \mathsf{SNR_2} +
\mathsf{INR_2})(1+ \mathsf{SNR_1} + \mathsf{INR_2})}{(\mathsf{1 +
INR_2}) (\mathsf{(1+SNR_1)(1+SNR_2)  + INR_2})} \right). \nonumber
\end{eqnarray}
Let $N \rightarrow 0$, the above rate pair asymptotically goes to
\begin{equation}
\left\{
  \begin{array}{l}
\D R_1 \approx \frac{1}{2} \log \left( 1 +
\frac{\mathsf{SNR_1}}{\mathsf{INR_2}} \right) + R_0\\
R_2 \approx \frac{1}{2} \log(\mathsf{SNR_2})
  \end{array}
\right.
\end{equation}
which again achieves the asymptotic sum capacity (\ref{sum_capacity_typeI}). We remark that this is akin to the capacity
result for a class deterministic relay channel \cite{Kim2008}, where
both decode-and-forward and compress-and-forward are shown to be capacity
achieving.

Although we have demonstrated the asymptotic sum-rate optimality of
the point $B$ and all points between $B$ and $C$ in the weak
interference regime as $N \rightarrow 0$ (while the ratios of SNRs and
INRs are kept fixed), we remark that the achievable region
(\ref{rate_region_Z_low}) may not be asymptotically optimal in other
regimes. For example, in the regime where $\mathsf{SNR}_2 \gg
\mathsf{INR}_2$, both the $R_1+R_2$ and $2R_1+R_2$ values at point $C$
($\beta=0$) are unbounded away from their corresponding upper bounds
as shown by Wang and Tse \cite[Lemma 5.1]{Wang_ReceiverCooperation}
(Eq. (22) and Eq.  (26)). To close this gap, one can use Wang and Tse's
quantize-map-and-forward approach \cite{Wang_ReceiverCooperation},
which in fact achieves the capacity region of the general Gaussian
interference channel with bidirectional links to within a constant
number of bits.

\section{Gaussian Z-Interference Channel with \\
	a Relay Link: Type II}

\subsection{Achievable Rate Region}

As a counterpart of the Type I channel considered in the previous
section, this section studies the Type II channel, where the relay
link goes from the interfered receiver to the interference-free
receiver as shown in Fig.~\ref{Gaussian_Z_introduction}(b).
Intuitively, when the interference link is weak, the digital link
would not be as efficient as in the Type I channel, because receiver
$1$'s knowledge of $X_2$ is inferior to that of the receiver $2$.
However, when the interference link is very strong, receiver $1$
becomes a better receiver for $X_2$ than receiver $2$, in which case
the digital link is capable of increasing user $2$'s rate by as
much as $R_0$. %This is illustrated in the achievability theorem below.
%This section makes these intuitions precise by deriving
%an achievability theorem for the weak and moderately strong
%interference regimes, and a capacity theorem for the strong and very
%strong interference regimes.

\begin{figure} [t]
\centering \psfrag{X1}{$X_1$} \psfrag{X2}{$X_2$} \psfrag{Y1}{$Y_1$}
\psfrag{Y2}{$Y_2$} \psfrag{U2}{$U_2$} \psfrag{W2}{$W_2$}
\psfrag{Z1}{$Z_1$} \psfrag{Z2}{$Z_2$} \psfrag{R0}{$R_0$}
\psfrag{S1}{$S_1$} \psfrag{T2}{$T_2$} \psfrag{S2}{$S_2$}
\psfrag{P1}{$P_1$} \psfrag{b_P2}{$\overline{\beta} P_2$}
\psfrag{bP2}{$\beta P_2$} \psfrag{h11}{$h_{11}$}
\psfrag{h22}{$h_{22}$} \psfrag{h21}{$h_{21}$}
\includegraphics[width=3.3in]{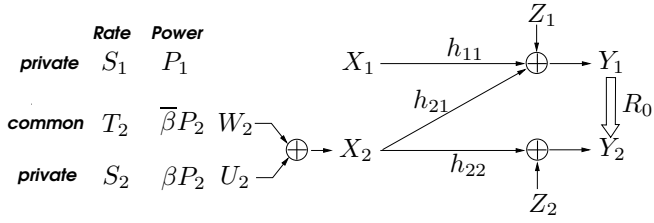}
\caption{Common-private power splitting for Type II channel with
$R_0 = R_a + R_b$, where $R_a$ is used to decode-and-forward $W_2$,
and $R_b$ is used to compress-and-forward $U_2$.}
\label{splitting_typeII}
\end{figure}

The main difference between the Type I and the Type II channels is that
in the Type I channel, the relay ($Y_2$) observes a noisy version of
the interference at the relay destination ($Y_1$). In addition, the
interference consists of messages intended for $Y_2$. Thus, the
decoding and the forwarding of the interference is a natural strategy.
In the Type II
channel, the relay ($Y_1$) observes a noisy version of the intended
signal at the relay destination ($Y_2$). Thus, decode-and-forward and
compress-and-forward can both be used.  The following achievability
theorem is based on a combination of the Han-Kobayashi scheme (with
$\beta$ being the common-private splitting ratio) and two relay
strategies, where the relay decodes then forwards the common
information using a rate $R_a$ and compresses then forwards the private
information using a rate $R_b$, with $R_a+R_b=R_0$, as shown in
Fig.~\ref{splitting_typeII}. In addition,
the presence of common information gives rise to the possibility of
compressing a combination of private and common messages.  A parameter
$\alpha$ accounts for the combination of private and common message
compression.

Unlike the Type I channel, the achievable rate region for the Type II
Gaussian Z-relay-interference channel has a more complicated
structure. In addition to the weak, strong and very strong
interference regimes, there is a new {\it{moderately strong}} regime,
where a combination of the decode-and-forward and the compress-and-forward
strategies is proposed. The proposed scheme reduces to pure
compress-and-forward in the weak interference regime, and pure
decode-and-forward in the strong interference regime.

%Before the theorem statement, we define the following
%$\delta(\beta, R)$ as a function of $\beta$ and $R$:
%\begin{eqnarray} \label{delta0}
%\delta_0(\beta, R) &=& \gamma \left( \frac{\beta
%(2^{2R}-1)\mathsf{INR_2(1+SNR_2)}}{\left( 2^{2R}(\mathsf{1 +SNR_2})
%+ \mathsf{INR_2}\right)(1 + \beta \mathsf{SNR_2})}\right)
%\end{eqnarray}

%\begin{figure} [t]
%\centering \psfrag{X1}{$X_1$} \psfrag{X2}{$X_2$} \psfrag{Z1}{$Z_1$}
%\psfrag{Z2}{$Z_2$} \psfrag{Y1}{$Y_1$} \psfrag{Y2}{$Y_2$}
%\psfrag{h11}{$h_{11}$}  \psfrag{h22}{$h_{22}$}
%\psfrag{h21}{$h_{21}$} \psfrag{R0}{$R_0$}
%\includegraphics[width=3.0in]{./figures/relay_ic_type_ii}
%\caption{Gaussian Z-interference channel with a relay link: type II.}
%\label{Z_relay_typeII}
%\end{figure}

\begin{theo} \label{theorem_Z_typeII}
For the Type II Gaussian Z-interference channel with a digital relay link
of limited rate $R_0$ from the interfered receiver to the interference-free
receiver as shown in Fig.~\ref{Gaussian_Z_introduction}(b), in the weak
interference regime defined by $\mathsf{INR_2 \le SNR_2}$, the
following rate region is achievable:
\begin{multline} \label{rate_region_weak_typeII}
\bigcup_{0 \leq \beta \leq 1}  \left \{ (R_1, R_2) \left| R_1 \leq
\gamma \left ( \frac{\mathsf{SNR}_1}{1 + \beta \mathsf{INR}_2}
\right), \right. \right. \\
R_2 \leq \gamma(\mathsf{\beta SNR_2}) +
\left. \gamma \left( \mathsf{\frac{\overline{\beta}INR_2}{1 + SNR_1
+ \beta INR_2}} \right) +  \delta(\beta, R_0) \right \},
\end{multline}
where
\begin{equation} \label{delta}
\delta(\beta, R_0) = \gamma \left( \frac{\beta
(2^{2R_0}-1)\mathsf{INR_2}}{2^{2R_0}(\mathsf{1 +\beta SNR_2}) +
\mathsf{\beta INR_2}}\right).
\end{equation}
In the moderately strong interference regime, defined by
\begin{equation} \label{condition_moderate_strong_typeII}
\mathsf{SNR_2 \le INR_2} \le 2^{2R_0}(1 + \mathsf{SNR_2}) -1
\stackrel{\bigtriangleup}{=} \mathsf{INR}_2^{\dagger},
\end{equation}
the following rate region is achievable:
\begin{equation} \label{rate_region_moderate_bin_typeII}
\mathrm{co}\left\{ \bigcup_{\alpha \in \mathbb{R},
	0 \le \beta \le 1, \;R_a+R_b \le R_0}
	\mathcal{R}_{\alpha,\beta}(R_a,R_b) \right \},
\end{equation}
where ``co" denotes convex hull and $\mathcal{R}_{\alpha,\beta}(R_a,R_b)$
is a pentagon region given by
\begin{equation} \label{R_beta_typeII}
\left\{ (R_1, R_2)\left|
\begin{array}{l}
R_1 \le \gamma (\mathsf{\frac{SNR_1}{1 + \beta INR_2}}) \\
R_2 \le \min \left\{ \gamma(\mathsf{SNR_2}) + R_b +
	\eta(\alpha,\beta,R_a), \right. \\
	\qquad \qquad \quad \gamma(\mathsf{\beta SNR_2}) + \gamma \left(
	\mathsf{\frac{\overline{\beta}INR_2}{1+\beta INR_2}} \right) \\
	\qquad \qquad \quad \left. + \zeta(\alpha,\beta,R_a) \right \} \\
R_1+ R_2 \le  \gamma(\beta \mathsf{SNR_2}) + \gamma \left(
	\mathsf{\frac{SNR_1 + \overline{\beta}INR_2}{1 + \beta INR_2}}\right) \\
	\qquad \qquad \quad + \zeta(\alpha,\beta, R_a)
\end{array}
\right. \right \},
\end{equation}
where
%\footnote{The definition of $\delta_2(\alpha,\beta,R_a)$ is
%provided for convenience and for completeness.}
\begin{equation}
\label{delta_1}
\zeta(\alpha,\beta,R_a) = \gamma \left( \frac{\beta \mathsf{INR_2}}
{(1+\beta \mathsf{SNR_2})(1+\frac{\sigma^2}{N})} \right),
\end{equation}
%\begin{equation}
%\label{delta_2}
%\delta_2(\alpha,\beta,R_a) = \gamma \left( \frac{\overline{\beta}
%	(1+\alpha)^2 \mathsf{INR_2}}
%{(1+ \overline{\beta}(1+\alpha)^2 \mathsf{SNR_2})(1+\frac{\sigma^2}{N})} \right),
%\end{equation}
and
\begin{multline}
\label{delta_12}
\eta(\alpha,\beta,R_a) = \\ \gamma \left(
\frac{ (1 + 2 \alpha \overline{\beta} + \alpha^2 \overline{\beta})
\mathsf{INR_2} +
\beta \overline{\beta} \alpha^2 \mathsf{INR_2} \mathsf{SNR_2} }
{(1+ \mathsf{SNR_2})(1+\frac{\sigma^2}{N})} \right)
\end{multline}
with
\begin{equation}
\label{sigma}
\frac{\sigma^2}{N} = \frac{ 1 + \mathsf{SNR_2} +
(1 + 2 \alpha \overline{\beta} + \alpha^2 \overline{\beta} )
\mathsf{INR_2} +
\beta \overline{\beta} \alpha^2 \mathsf{INR_2} \mathsf{SNR_2} }
{ (2^{2R_a}-1)(1+\mathsf{SNR_2}) }.
\end{equation}
In the strong interference regime defined by
\begin{equation} \label{condition_strong_typeII}
\mathsf{INR}_2^{\dagger} \le \mathsf{INR_2} \le \mathsf{(1 + SNR_1)}
\mathsf{INR}_2^{\dagger} \stackrel{\bigtriangleup}{=}
\mathsf{INR}_2^{\ddag},
\end{equation}
the capacity region is given by
\begin{equation} \label{capacity_region_strong_Z_typeII}
\left\{ (R_1, R_2)\left|
  \begin{array}{rll}
R_1 &\le& \gamma (\mathsf{SNR_1})\\
R_2 &\le& \gamma (\mathsf{SNR_2}) +R_0  \\
R_1+R_2 &\le& \gamma(\mathsf{SNR_1 + INR_2})
\end{array}
\right. \right \}.
\end{equation}
In the very strong interference regime defined by
\begin{equation} \label{condition_strong_Z_typeII}
\mathsf{INR_2} \ge \mathsf{INR}_2^{\ddag},
\end{equation}
the capacity region is given by
\begin{equation} \label{capacity_region_very_strong_Z_typeII}
\left\{ (R_1, R_2)\left|
  \begin{array}{l}
 R_1 \le \gamma(\mathsf{SNR_1})\\
R_2 \le \gamma (\mathsf{SNR_2}) + R_0
  \end{array}
\right. \right \}.
\end{equation}
\end{theo}

\begin{proof}
See Appendix \ref{appendix_theorem_3}.
\end{proof}

%\subsection{Comments on Theorem~\ref{theorem_Z_typeII}}
\subsection{Numerical Examples}

\begin{figure} [t]
\centering
\includegraphics[width=3.5in]{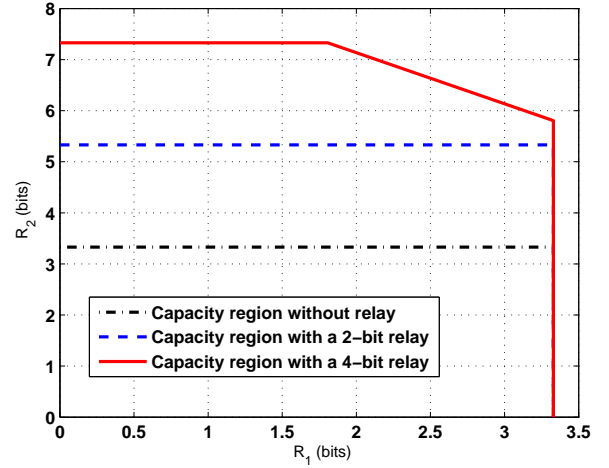}
\caption{Achievable region of the Gaussian Z-interference channel in
the strong interference regime with and without a digital relay
link of Type II.} \label{pic_strong_capacity_region_typeII}
\end{figure}

In the strong and very strong interference regimes, the entire $X_2$
is common information.  The relay expands the capacity region by
decoding $X_2^n$ at receiver $1$ and forwarding its bin index to receiver $2$.  The boundaries of the strong and very strong regimes depend on the relay
link rate.
%Due to the strong
%interference link, this common message $X_2$ is guaranteed to be
%decodable at $Y_1$.
As a numerical example, Fig.~\ref{pic_strong_capacity_region_typeII}
shows how the capacity region of a Type II channel is expanded by the
relay link in the strong and very strong interference regimes.  Here,
$\mathsf{SNR_1=SNR_2=20}$ dB and $\mathsf{INR_2=55}$ dB. Without the
digital link, this is a Gaussian Z-interference channel in the
very strong interference regime \cite{Sato}, where $\mathsf{INR_2 \ge
SNR_2(1+SNR_1)}$ and the capacity region is a rectangle as depicted by
the dash-dotted region in
Fig.~\ref{pic_strong_capacity_region_typeII}. With a 2-bit digital
link, $R_2$ is expanded by exactly 2 bits. The Z-interference channel
remains in the very strong interference regime, where the capacity
region is given by (\ref{capacity_region_very_strong_Z_typeII}) and
depicted by the dashed rectangular region in
Fig.~\ref{pic_strong_capacity_region_typeII}.
When $R_0=4$ bits, the Z-interference channel now falls into
the strong interference regime.  The capacity region as given by
(\ref{capacity_region_strong_Z_typeII}) now becomes a solid pentagon
region.
%In this regime, for some $R_1$ lower than a certain threshold,
%the rate increase in $R_2$ is exactly 4 bits. For other values of $R_1$,
%the rate increase in $R_2$ is less than 4 bits.
Further increase in the rate of the digital link can increase the
maximum $R_2$ but not the sum rate.

\begin{figure} [t]
\centering
\begin{overpic}[scale=0.6]{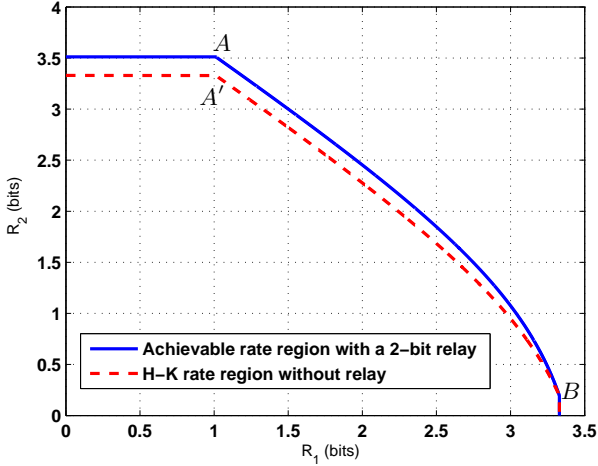}
\centering
\put(35,63){\small $A$}
\put(33,55){\small $A'$}
\put(87,11){\small $B$}
\end{overpic}
\caption{Capacity region of the Gaussian Z-interference channel in
the weak interference regime with and without a digital relay link
of Type II.} \label{pic_weak_region_typeII}
\end{figure}

In the weak interference regime where $\mathsf{INR_2 \le SNR_2}$,
Theorem~\ref{theorem_Z_typeII} shows that a pure
compress-and-forward for the private message should be used for relaying.
Intuitively, this is because when the interference link is weak
the common message rate is limited by the interference link, which
cannot be helped by relaying. Thus, the digital link needs to focus on
helping the decoding of private message at $Y_2$ by
compress-and-forward.
As a numerical example, Fig.~\ref{pic_weak_region_typeII} shows the
achievable rate region of a Gaussian Z-interference channel with
$\mathsf{SNR_1=SNR_2=20}$ dB and $\mathsf{INR_2=15}$ dB with and
without the relay link. The dashed region denoted by
points $A'$ and $B$ represents the rate region achieved without the
digital link. The solid rate region denoted by points $A$ and $B$ is
with a 2-bit digital link.
%In Fig.~\ref{pic_weak_region_typeII}, points $A$ and $A'$ both
%correspond to $\beta=1$, where the entire $X_2$ is the private
%message.  As $\beta$ decreases from 1 to 0, the rate pair moves from
%point $A$ (or $A'$) to point $B$, which corresponds to $\beta=0$,
%where the entire $X_2$ is common message.
From the rate pair
expression (\ref{rate_region_weak_typeII}), the effect of the digital
link is to shift the rate region of the channel without the relay
upward by $\delta(\beta, R_0)$ bits. Since $\delta(\beta, R_0)$ is
monotonically decreasing as $\beta$ decreases from 1 to 0, for fixed
$R_1$, the largest increase in $R_2$ corresponds to $\delta(1, R_0)$,
i.e. the increase from point $A'$ to $A$.  Note that $A$ and $A'$
are the maximum sum-rate points of the Type II Gaussian Z-interference
channel with and without the relay respectively.  They correspond to all-private
message transmission, which is in contrast to the Type I case where
the maximum sum rate is achieved with some $\beta^* \neq 1$. Finally,
we note that the relay does not affect point $B$, which corresponds to
$\beta=0$, because $\delta(0,R_0)=0$.

\subsection{Sum-Capacity Upper Bound}

%For the Type I channel, the relay link asymptotically
%achieves the sum capacity cut-set bound in the weak interference
%regime and high SNR/INR limit. For the Type II channel, however,
%the cut-set bound is not achievable.
%Recall that, we proved in Theorem~\ref{theorem_Z_typeII} that in the
%weak interference regime, the rate pair in
%(\ref{rate_region_weak_typeII}) is the lower-right corner point of
%the pentagon defined by $f_1(\beta)$, $f_2(\beta)$ and $f_3(\beta)$,
%and $f_3(\beta)$ is a monotonic increasing function of $\beta$ when
%$\mathsf{INR_2 \le SNR_2}$. Therefore,

By Theorem~\ref{theorem_Z_typeII}, an achievable sum rate of
the Type II Gaussian Z-interference channel with a relay link
of rate $R_0$ in the weak interference regime is
\begin{equation}
\label{R_sum_comp_1_II}
R_{sum} = \gamma \left( \frac{\mathsf{SNR}_1}{1 + \mathsf{INR}_2}
\right) + \gamma(\mathsf{SNR_2}) + \delta(1, R_0),
\end{equation}
which is obtained by setting $\beta=1$ in
(\ref{rate_region_weak_typeII}).
%and adding up $R_1$ and $R_2$.
%Applying the high SNR/INR extreme condition
%(\ref{extreme_condition}), the asymptotic sum rate is obtained as
%\begin{eqnarray}
%R_{sum} &\approx& \frac{1}{2} \log \left
%(\mathsf{\frac{SNR_2(SNR_1+INR_2)}{INR_2}} \right) +  \nonumber \\
%&&\qquad  \frac{1}{2}  \log \left(  \frac{\mathsf{SNR_2
%+INR_2}}{\mathsf{SNR_2} + 2^{-2R_0}\mathsf{INR_2}} \right) \nonumber \\
%&\approx& C_{sum}(0) + \frac{1}{2}  \log \left(  \frac{\mathsf{SNR_2
%+INR_2}}{\mathsf{SNR_2} + 2^{-2R_0}\mathsf{INR_2}} \right) \nonumber
%\end{eqnarray}
Comparing with the sum capacity of the Gaussian Z-interference channel
without the relay in the weak interference regime (\ref{define_C_sum0}),
the sum-rate increase using the relay scheme of
Theorem~\ref{theorem_Z_typeII} is upper bounded by
\begin{eqnarray}
\delta(1, R_0)
&=& \frac{1}{2} \log \left( \frac{\mathsf{1 + SNR_2
+INR_2}}{\mathsf{1 + SNR_2} + 2^{-2R_0}\mathsf{INR_2}} \right)
\nonumber\\
&\le & \gamma\left( \frac{\mathsf{INR_2}}{\mathsf{1+SNR_2}}\right)
\nonumber\\
&\le & \frac{1}{2},
\label{R_sum_comp_2_II}
\end{eqnarray}
%which is always smaller than the cut-set bound
%$R_0$ in the weak interference regime, where $\mathsf{INR_2 \le
%SNR_2}$. In fact this sum-rate increase is upper bounded by 0.5 bit.
%Since $\triangle(R_0)$ is an increasing function of $R_0$, taking
%the limit on $\triangle(R_0)$, we have
%\begin{equation}
%\lim_{R_0 \rightarrow \infty} \triangle(R_0) = \gamma\left(
%\mathsf{\frac{INR_2}{SNR_2}} \right)  \le \frac{1}{2}
%\end{equation}
where $\mathsf{INR_2 \le SNR_2}$ is used in the last step. As
illustrated in the example in Fig.~\ref{pic_weak_region_typeII},
the rate increase from point $A'$ to point $A$ is about 0.2
bits, which is less than 1/2 bits and is a fraction of the 2-bit
relay link rate.  This is in contrast to the Type I
channel, where each relay bit can increase the
sum rate by up to one bit.  The following theorem provides an
asymptotic sum-capacity result for the Type II channel and a proof of
the 1/2-bit upper bound when $\mathsf{INR_2}$ is not very strong.

%\begin{theo} \label{theorem_sum_capacity_typeII}
%Let $\mathcal{C}(R_0)$ denote the capacity region of the Type II
%Gaussian Z-interference channel with a digital realy link of limited
%rate $R_0$ from the interfered receiver to the interference-free
%receiver as shown in Fig.~\ref{Gaussian_Z_introduction}(b). Let
%\begin{equation}
%C_{sum}^\theta(R_0) = \max_{(C_1,C_2) \in \mathcal{C}(R_0)}
%	\theta C_1 + (1-\theta) C_2.
%\end{equation}
%Consider the weak interference regime, where $\mathsf{INR_2} \le
%\mathsf{SNR_2}$. For any $R_0 \ge 0$, a capacity upper bound
%for the Gaussian Z-relay-interference channel is
%\begin{equation} \label{sum_capacity_typeII}
%C^\theta_{sum}(R_0) \le C^\theta_{sum}(0) +
% (1-\theta) \gamma\left(\frac{\mathsf{INR_2}}{\mathsf{1+SNR_2}}\right).
%\end{equation}
%In particular, as $R_0 \rightarrow \infty$, the asymptotic sum-capacity
%of the Type II Gaussian Z-interference channel with a relay link is
%\begin{equation}
%\label{degraded_sum_capacity}
%\gamma \left( \frac{\mathsf{SNR}_1}{1 + \mathsf{INR}_2} \right) +
%\gamma\left(1+\mathsf{INR_2}+{\mathsf{SNR_2}}\right).
%%\gamma(\mathsf{SNR_2}) +
%%\gamma\left(\frac{\mathsf{INR_2}}{\mathsf{1+SNR_2}}\right).
%\end{equation}
%Finally, the sum capacity of the Type II Gaussian Z-relay-interference
%channel is upper bounded by the sum capacity of the Gaussian Z-interference
%channel without a relay link plus half a bit.
%\end{theo}
\begin{theo} \label{theorem_sum_capacity_typeII}
For the Type II Gaussian Z-interference channel with a digital relay
link of rate $R_0$ from the interfered receiver to the
interference-free receiver as shown in
Fig.~\ref{Gaussian_Z_introduction}(b), when $R_0 \rightarrow \infty$,
the asymptotic sum capacity is
\begin{equation} \label{degraded_sum_capacity}
C_{sum}(\infty) = \gamma(\mathsf{SNR_1 + INR_2}) + \gamma \left ( \mathsf{\frac{SNR_2}{1+INR_2}}\right).
\end{equation}
Further, when $\mathsf{INR_2 \le INR_2^{\S}}$, where $\mathsf{INR_2^{\S}}$
is defined by $\mathsf{INR_2^{\S} = SNR_2(1+SINR_1)}$, we have
\begin{equation}
C_{sum}(\infty) - C_{sum}(0) \le \frac{1}{2}.
\end{equation}
%and when $\mathsf{INR_2 \ge INR_2^{+}}$ and $\min\{\mathsf{INR_2, SNR_1, SNR_2}\} \gg 1$, $C_{sum}(\infty) - C_{sum}(0) \rightarrow \infty$ as
%$\mathsf{\frac{INR_2}{SNR_1 SNR_2}} \rightarrow \infty$.
\end{theo}

\begin{proof}
When $R_0 = \infty$, receiver $2$ has complete knowledge of $Y_1^n$.
Starting from Fano's inequality:
\begin{eqnarray} \label{type_2_fano's bound}
\lefteqn{n(R_1 + R_2)} \nonumber \\
& \le & I(X_1^n; Y_1^n) + I(X_2^n; Y_1^n, Y_2^n) + n \epsilon_n \nonumber \\
& \stackrel{(a)}{\le} & I(X_1^n; Y_1^n) + I(X_2^n; Y_1^n, Y_2^n | X_1^n) + n \epsilon_n \nonumber \\
%& = & I(X_1^n; Y_1^n) + I(X_2^n; Y_1^n |X_1^n) + I(X_2^n; Y_2^n|Y_1^n, X_1^n)  + n \epsilon_n \nonumber \\
& = & I(X_1^n, X_2^n; Y_1^n) + I(X_2^n; Y_2^n|Y_1^n, X_1^n)  + n \epsilon_n,
\end{eqnarray}
where (a) follows from the fact that $X_1^n$ is independent of $X_2^n$.
%in which case
%\begin{eqnarray}
%I(X_2^n; Y_1^n, Y_2^n) &=& h(X_2^n) -h(X_2^n | Y_1^n, Y_2^n) \nonumber \\
%&=& h(X_2^n|X_1^n) -h(X_2^n | Y_1^n, Y_2^n) \nonumber \\
%&\le& h(X_2^n|X_1^n) -h(X_2^n | Y_1^n, Y_2^n, X_1^n) \nonumber \\
%&=& I(X_2^n; Y_1^n, Y_2^n|X_1^n)
%\end{eqnarray}
%Clearly, $X_1$, $X_2$ and $Y_1$ forms a multiple-access channel.
%As a result, the first term of (\ref{type_2_fano's bound}) is bounded by
The first term in (\ref{type_2_fano's bound}) is bounded by the sum
capacity of the multiple-access channel $(X_1^n,X_2^n,Y_1^n)$:
\begin{equation} \label{type_2_bound_1}
I(X_1^n, X_2^n; Y_1^n) \le n \gamma (\mathsf{SNR_1+INR_2})
\end{equation}
%with equality achieved by independent Gaussian codebooks.
The second term in (\ref{type_2_fano's bound}) is bounded by
\begin{eqnarray} \label{type_2_bound_2}
\lefteqn{I(X_2^n; Y_2^n|Y_1^n, X_1^n)} \nonumber \\
&=& h(Y_2^n|Y_1^n, X_1^n) - h(Y_2^n|Y_1^n, X_1^n, X_2^n) \nonumber \\
&\stackrel{(a)}{\le}& \sum_{i=1}^{n} \left \{h(Y_{2,i} | Y_{1,i}, X_{1,i}) -h(Z_{2,i}) \right \} \nonumber \\
&=& \sum_{i=1}^{n} \left \{ h(h_{22}X_{2,i} +Z_{2,i}| h_{21}X_{2,i}+Z_{1,i}) -h(Z_{2,i}) \right \} \nonumber \\
&\stackrel{(b)}{\le}& n \gamma \left( \mathsf{\frac{SNR_2}{1+INR_2}}\right),
\end{eqnarray}
where (a) follows from the chain rule and the fact that conditioning does not increase entropy, and (b) follows from the fact that Gaussian distribution maximizes the conditional entropy under a covariance
constraint.
%\begin{eqnarray}
%h(Y_2^n|Y_1^n, X_1^n, X_2^n) = h(Z_2^n) &=& \sum_{i=1}^{n} h(Z_2^i) \nonumber \\
%h(Y_2^n|Y_1^n, X_1^n) \le \sum_{i=1}^{n}h(Y_2^i|Y_1^n, X_1^n) &\le& \sum_{i=1}^{n}h(Y_2^i|Y_1^i, X_1^i) \nonumber
%\end{eqnarray}
Combining (\ref{type_2_bound_1}) and (\ref{type_2_bound_2}) gives the
sum rate upper bound:
\begin{equation} \label{type_2_upper_bound}
C_{sum}(\infty) \le \gamma(\mathsf{SNR_1 + INR_2}) + \gamma \left ( \mathsf{\frac{SNR_2}{1+INR_2}}\right).
\end{equation}
It can be easily verified that the above sum-rate upper bound is also
asymptotically achievable. By Theorem~\ref{theorem_Z_typeII}, with
$R_0 =\infty$, there are only two interference regimes: weak interference regime and moderately strong interference regime. In the weak
interference regime, a pure compress-and-forward scheme, i.e., setting
$\beta =1$ in (\ref{rate_region_weak_typeII}) achieves
(\ref{type_2_upper_bound}). In the moderately strong interference regime,
setting $\beta = 1$ and $R_b = 0$ in (\ref{R_beta_typeII}) achieves
\begin{equation}
\gamma(\mathsf{SNR_2}) + \gamma \left ( \mathsf{\frac{SNR_1}{1+INR_2}}\right)
+ \gamma \left ( \mathsf{\frac{INR_2}{1+SNR_2}}\right)
\end{equation}
which is equivalent to (\ref{type_2_upper_bound}). This proves the asymptotic sum-capacity result.

Now, without the relay link, the sum capacity for the Gaussian Z-interference
channel is (\cite{Carleial, Sato, HK1981,Sason2004, Tse2007}):
\begin{eqnarray*} \label{sum_capacity_without_relay}
\lefteqn{C_{sum}(0) =} \nonumber \\
&
\left\{
  \begin{array}{ll}
	\gamma(\mathsf{SNR_2}) +
	\D \gamma \left ( \mathsf{\frac{SNR_1}{1+INR_2} }\right)
		& \mathrm{if\ \ } \mathsf{INR_2 \le SNR_2}  \\
	\gamma(\mathsf{SNR_1+INR_2})
		& \mathrm{if\ \ } \mathsf{SNR_2 \le INR_2 \le INR_2^{\S}} \\
	\gamma(\mathsf{SNR_1}) + \gamma(\mathsf{SNR_2})
		& \mathrm{if\ \ } \mathsf{INR_2 \ge INR_2^{\S}}
  \end{array}
\right.
&
\end{eqnarray*}
Comparing
%(\ref{sum_capacity_without_relay})
$C_{sum}(0)$ with the asymptotic sum capacity in
the limit of large relay rate (\ref{degraded_sum_capacity}), we have
\begin{equation}
C_{sum}(\infty) - C(0) = \gamma \left(\mathsf{\frac{INR_2}{1+SNR_2}}  \right) \le \frac{1}{2}
\end{equation}
when $\mathsf{INR_2 \le SNR_2}$ and
\begin{equation}
C_{sum}(\infty) - C(0) = \gamma \left(\mathsf{\frac{SNR_2}{1+INR_2}}  \right) \le \frac{1}{2}
\end{equation}
when $\mathsf{SNR_2 \le INR_2 \le INR_2^\S}$. Therefore, the sum-capacity
gain is upper bounded by half a bit when $\mathsf{INR_2 \le INR_2^{\S}}$.
\end{proof}

Note that when $\mathsf{INR_2 \ge INR_2^\S}$, the sum-capacity gain
can be larger than half a bit. In fact,
in the regime where $\mathsf{INR_2 \gg SNR_1, INR_2 \gg SNR_2}$
and $\mathsf{SNR_1, SNR_2} \gg 1$, we have
\begin{equation}
C_{sum}(\infty) - C_{sum}(0) \approx
\frac{1}{2} \log \left( \mathsf{\frac{INR_2}{SNR_1 SNR_2}} \right),
\end{equation}
which can be unbounded.

The asymptotic sum capacity (\ref{degraded_sum_capacity}) is
essentially the sum capacity of a degraded Gaussian interference
channel where the inputs are $X_1$ and $X_2$, and outputs are $Y_1$
and $(Y_1,Y_2)$ of a Gaussian Z-interference channel. The capacity
region for the general degraded interference channel is still open.

\section{Summary}

This paper studies a Gaussian Z-interference channel with
unidirectional relay link at the receiver.
%, where a rate-limited digital link is provided
%from one receiver to the other one.
When the relay link goes from the interference-free receiver
to the interfered receiver, a suitable relay strategy is to
let the interference-free receiver decode-and-forward a part
of the interference for subtraction. Interference
decode-and-forward is capacity achieving in the strong interference
regime.  In the weak interference regime, the asymptotic sum capacity can be achieved with either a decode-and-forward
or a compress-and-forward strategy in the high SNR/INR limit.

When the relay link goes from the interfered receiver to the
interference-free receiver, a suitable relay strategy is a combination
of decode- and compress-and-forward of the intended message. In the
strong interference regime, decode-and-forward alone is capacity
achieving.  In the weak interference regime, the combination scheme
reduces to pure compress-and-forward. In the
moderately strong interference regime, a combination of both need
to be used.

The direction of the relay link is crucial. In the weak interference
regime, a relay link from the interference-free receiver to the
interfered receiver can significantly increase the achievable sum rate
by up to one bit for every relay bit, while in the reversed direction,
the sum rate increase is upper bounded by half a bit regardless of the
relay link rate. In contrast, in the strong interference regime, the
sum-capacity gain due to a relay from the interference-free receiver
to the interfered receiver eventually saturates, while a relay link in the reverse direction provides unbounded sum-capacity gain.

\appendix

\subsection{Convexity of Achievable Rate Region (\ref{convex_region_chapter_2})}
\label{appendix_convexity}

This appendix shows that the region defined by $R_1 \le
\mathsf{\gamma(SNR_1)}$, $R_2 \le \mathsf{\gamma(SNR_2)}+R_0$, and
the curve
\begin{equation} \label{concave_curve}
\left\{
\begin{array}{lll}
R_1 & = & \gamma \left ( \D \frac{\mathsf{SNR}_1}{1 +
    \beta \mathsf{INR}_2} \right) \\
R_2 & = & \gamma(\beta \mathsf{SNR}_2) +
    \gamma \left(\D \frac{\overline{\beta}\mathsf{INR}_2}{1 +
    \mathsf{SNR}_1 + \beta \mathsf{INR}_2} \right) +R_0
\end{array}
\right.
\end{equation}
where $0 \le \beta \le 1$, is convex when $\mathsf{INR_2 \le
SNR_2}$.

\begin{figure} [t]
\centering \psfrag{R1}{$R_1$}  \psfrag{R2}{$R_2$} \psfrag{0}{$0$}
\psfrag{A}{$\gamma(\mathsf{\frac{SNR_1}{1+INR_2}})$}
\psfrag{C}{$\mathsf{\gamma(SNR_1)}$}
\psfrag{D}{$\mathsf{\gamma(\frac{INR_2}{1+SNR_1})}$}
\psfrag{B}{$\mathsf{\gamma(SNR_2)} +R_0$}
\psfrag{P}{\small $A$} \psfrag{Q}{\small $B$}
\includegraphics[width=2.5in]{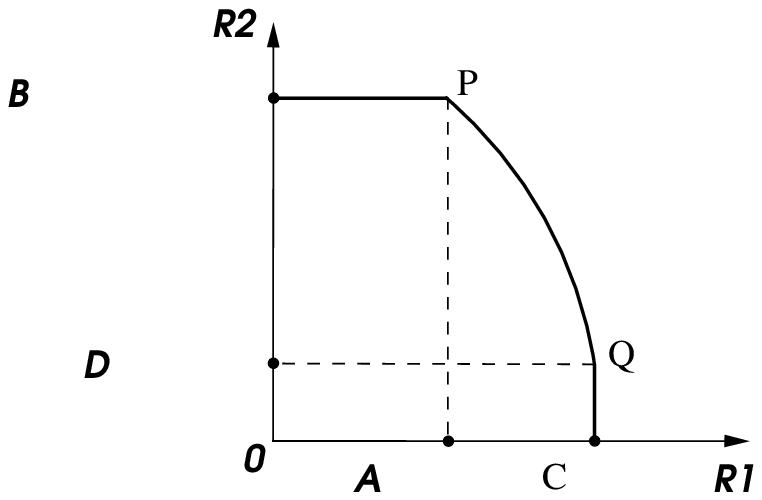}
\caption{The region defined by lines  $R_1 =
\mathsf{\gamma(SNR_1)}$, $R_2 = \mathsf{\gamma(SNR_2)}+R_0$ and the
curve (\ref{concave_curve}).} \label{convex_region}
\end{figure}

Note that, when $\beta=1$ and $\beta=0$, the curve defined by
(\ref{concave_curve}) meets $R_2 = \mathsf{\gamma(SNR_2)}+R_0$ and
$R_1 = \mathsf{\gamma(SNR_1)}$ at points $A$ and $B$, respectively, as shown in
Fig.~\ref{convex_region}. Therefore, to prove the convexity of the
region, we only need to prove that the curve (\ref{concave_curve})
parameterized by $\beta$ is concave.
%\begin{equation}
%\left\{
%\begin{array}{l}
%R_1^{P}  =  \gamma \left ( \D \mathsf{\frac{SNR_1}{1+INR_2}} \right) \\
%R_2^{P}  =  \gamma \left(\D \mathsf{SNR_2}\right) +R_0
%\end{array}
%\right. \quad \left\{
%\begin{array}{l}
%R_1^{Q}  =  \gamma \left(\D \mathsf{SNR_1}\right) \\
%R_2^{Q}  =  \gamma \left ( \D \mathsf{\frac{INR_2}{1+INR_1}} \right)
%+ R_0
%\end{array}
%\right.
%\end{equation}

First, we express $\beta$ in terms of $R_1$:
\begin{equation} \label{beta_as_function_of_R1}
\beta = \frac{1}{\mathsf{INR_2}}\left(
\frac{\mathsf{SNR_1}}{2^{2R_1}-1} -1 \right).
\end{equation}
Substituting this expression for $\beta$ into the expression for
$R_2$ in (\ref{concave_curve}), we obtain $R_2$ as a function of
$R_1$:
\begin{equation} \label{R2_as_a_function_of_R1}
R_2 = \frac{1}{2} \log \left( -\nu 2^{2R_1} + \lambda \right) +
\mu
\end{equation}
where $\nu = \mathsf{\frac{SNR_2}{INR_2}} -1$,
$\lambda = \mathsf{\frac{SNR_2}{INR_2}(1+SNR_1)} -1$ and
$\mu = \gamma \left( \mathsf{\frac{1+INR_2}{SNR_1}} \right) +R_0$.
Note that when $\mathsf{INR_2 \le SNR_2}$, $\nu \ge 0$ and
$\lambda > 0$.

Observe that $R_1$ is a monotonic decreasing function of $\beta$.
So, in the range $0 \le \beta \le 1$, we have
\begin{equation} \gamma \left(
\mathsf{\frac{SNR_1}{1+INR_2}} \right )  \le R_1 \le
\gamma(\mathsf{SNR_1}).
\end{equation}
In this range of $R_1$, it is easy to verify that $-\nu 2^{2R_1}
+ \lambda > 0$.

Now, taking the first and second order derivatives of $R_2$ with
respect to $R_1$ in (\ref{R2_as_a_function_of_R1}), we have
\begin{eqnarray}
R_2' = \frac{-\nu 2^{2R_1}}{-\nu 2^{2R_1} + \lambda}, \;\;\;\; R_2'' = \frac{-2\lambda \nu 2^{2R_1}}{(-\nu 2^{2R_1} +
\lambda)^2} \ln 2.
\end{eqnarray}
Since $\nu \ge 0$, $\lambda > 0$, and $-\nu 2^{2R_1} + \lambda
> 0$, we have $R_2' \le 0$ and $R_2'' \le 0$. As a result, the curve
(\ref{concave_curve}) parameterized by $\beta$ is concave.

\subsection{Converse Proof for the Strong and Very Strong Interference
Regimes in Theorem \ref{theorem_z_type_I}}
\label{appendix_strong_I}

In this appendix, we prove a converse in the strong and very strong
interference regimes for the Type I channel. The converse is based on
a technique used in \cite{HK1981} and \cite{Sato} for proving the
converse for the strong interference channel without the relay link.
The idea is to show that when $\mathsf{INR_2 \ge \min \{SNR_2, INR_2^*\}}$,
if a rate pair $(R_1,R_2)$ is achievable for the Gaussian
Z-interference channel with a relay link, i.e., $X_1^n$ can be reliably
decoded at receiver $1$ at rate $R_1$, and $X_2^n$ can be reliably
decoded at receiver $2$ at rate $R_2$, then $X_2^n$ must also be decodable
at the receiver $1$.

First, the reliable decoding of $X_2^n$ at receiver $2$ requires
\begin{equation} \label{R_2_direct_constraint}
R_2 \le \mathsf{\gamma(SNR_2)}.
\end{equation}
To show that $X_2^n$ is also decodable at receiver $1$ when $\mathsf{INR_2 \ge
\min \{SNR_2, INR_2^*\}}$, consider the two cases
$\mathsf{SNR_2 \le INR_2^*}$ and $\mathsf{SNR_2 \ge INR_2^*}$
separately.

First, when $\mathsf{SNR_2 \le INR_2^*}$, we have $\mathsf{INR_2 \ge
SNR_2}$, or $h_{21} \ge h_{22}$.  In this case, after $X_1^n$ is decoded
at receiver $1$ (possibly with the help of the relay link), receiver $1$ may
subtract $X_1^n$ from $Y_1^n$ then scale the resulting signal to obtain
\begin{equation}
Y_1^{'n}= \frac{h_{22}}{h_{21}}(Y_1^n - h_{11}X_1^n) = h_{22}X_2^n
+\frac{h_{22}}{h_{21}}Z_1^n.
\end{equation}
When $h_{21} \ge h_{22}$, the
Gaussian noise $\frac{h_{22}}{h_{21}}Z_1^n$ in this effective channel has a smaller variance than the noise in $Y_2^n = h_{22} X_2^n + Z_2^n$.
Since $X_2^n$ is reliably decodable at receiver $2$, $X_2^n$ must also be reliably
decodable at receiver $1$.

When $\mathsf{SNR_2 \ge INR_2^*}$, we have $\mathsf{INR_2 \ge INR_2^*}$.
In this case, since $X_2^n$ is reliably decoded at $Y_2$, with the perfect
knowledge of $X_2^n$ at receiver $2$, $(X_2^n,Y_1^n,Y_2^n)$ forms a
deterministic relay channel \cite{Kim2008} with $X_2^n$ as the input,
$Y_1^n$ as the output and $Y_2^n$ as the deterministic relay. As a result,
the following rate for $X_2^n$ can be supported:
\begin{equation}
R_2 = \gamma \left(\mathsf{\frac{INR_2}{1+SNR_1}}\right ) + R_0
\end{equation}
Since $\mathsf{INR_2 \ge INR_2^*}$, it is easy to verify that the
above rate is always greater than the rate supported at the receiver $2$,
i.e.,
\begin{equation}
 \gamma \left(\mathsf{\frac{INR_2}{1+SNR_1}}\right ) + R_0   \ge
\gamma(\mathsf{SNR_2}),
\end{equation}
which implies that whenever $X_2^n$ is reliably decodable at $Y_2$,
it is also reliably decodable at $Y_1$ with the help of the relay.

Now, because both $X_1^n$ and $X_2^n$ are always decodable at receiver $1$ in
the strong interference regime, the achievable rate region of the Gaussian
Z-interference channel with a digital relay link is included in the
capacity region of the same channel in which both $X_1^n$ and $X_2^n$ are
required at $Y_1^n$, and $X_2^n$ is required at $Y_2^n$. Further, the
capacity region of such a channel can only be enlarged if $X_2^n$ is
provided to $Y_2^n$ by a genie. In such a case, the channel reduces to a
Gaussian multiple-access channel with $(X_1^n, X_2^n)$ as inputs, $Y_1^n$
as the output, and with the same relay
link from receiver $2$ to receiver $1$, where the relay knows $X_2^n$ perfectly. The capacity
region of such a channel is
\begin{equation} \label{mac_relay_region}
\left\{ (R_1, R_2) \left|
  \begin{array}{rll}
R_1 & \le&  \gamma (\mathsf{SNR_1}) \\
R_2 & \le&  \gamma (\mathsf{INR_2}) +R_0 \\
R_1+R_2 & \le&  \gamma(\mathsf{SNR_1 + INR_2}) +R_0
  \end{array} \right.
\right \}
\end{equation}
Combining (\ref{mac_relay_region}) and (\ref{R_2_direct_constraint}),
then applying (\ref{INRstarcondition}) gives us
(\ref{capacity_region_Z_moderate}). This proves that when
$\mathsf{INR_2 \ge \min\{SNR_2, INR_2^*\}}$, the
achievable rate region of the Gaussian Z-interference channel with a
relay link must be included in (\ref{capacity_region_Z_moderate}),
which, in the very strong interference regime, reduces to
(\ref{capacity_region_Z_strong}).
%This concludes the proof of the capacity region results in the
%strong and very strong interference regimes.

\subsection{Evaluation of Wyner-Ziv Rate (\ref{rate_pair_quantization})}
\label{appendix_wyner_ziv}

In this appendix, we show that with independent Gaussian inputs $X_1
\sim \mathcal{N}(0, P_1)$ and $X_2 \sim \mathcal{N}(0, P_2)$, and
the Gaussian quantization scheme
(\ref{Gaussian_noise_quantization}), the achievable rate described
by (\ref{R_2_quantization}), (\ref{R_1_quantization}) and
(\ref{Wyner-Ziv_constraint}) is given by
(\ref{rate_pair_quantization}). The technique is similar to that in
\cite{Maric_oneshot}.

With a Gaussian input $X_2 \sim \mathcal{N}(0, P_2)$, $R_2$ is given by
\begin{eqnarray}
R_2 &=& I(X_2; Y_2) = \gamma(\mathsf{SNR_2}).
\end{eqnarray}
With the knowledge of $X_2$ at $Y_2$, $X_1$, $Y_1$ together with $Y_2$ become
a deterministic relay channel with a digital link.
To fully utilize the digital link, we set $\hat{Y}_2$ to
be such that $I(Y_2; \hat{Y}_2 | Y_1) = R_0$. Note that $\hat{Y}_2= Y_2 +e$,
where $Y_2$ and $e$ are independent and $e \sim \mathcal{N}(0,\sigma^2)$.
To find $\sigma^2$, note that
\begin{eqnarray}
R_0 = h(\hat{Y}_2|Y_1) -h(\hat{Y}_2|Y_1, Y_2) = \gamma \left( \frac{\sigma_{Y_2|Y_1}^2}{\sigma^2}\right)
 \label{sigma_and_Y_2_Y_1}
\end{eqnarray}
where
$\sigma_{Y_2|Y_1}^2$, the conditional variance of
$Y_2$ given $Y_1$, can be calculated in a standard way. Thus, from 
(\ref{sigma_and_Y_2_Y_1}), we have
\begin{eqnarray}
\sigma^2 = \frac{N}{2^{2R_0}-1}\left( 1 +
\mathsf{\frac{SNR_2(1+SNR_1)}{1+SNR_1+INR_2}} \right).
\label{sigma_quantize_typeI}
\end{eqnarray}
%This gives us the distribution of $e \sim \mathcal{N}(0,\sigma^2)$.

Now, we are ready to calculate $R_1$. First,

%\begin{eqnarray}
%\lefteqn{h(\hat{Y}_2 | Y_1, X_1)} \nonumber \\
%&=& h(h_{22}X_2 + Z_2 + e | h_{11}X_1 + h_{21}X_2 + Z_1, X_1)
%\nonumber \\
%&=& h(h_{22}X_2 + Z_2 + e | h_{21}X_2 + Z_1)
%\nonumber \\
%&=& \frac{1}{2}\log  \left( 2\pi e\left( \sigma^2 + N +
%\frac{|h_{22}|^2P_2N}{|h_{21}|^2P_2 + N} \right) \right) \nonumber \\
%&=& \frac{1}{2} \log  \left(2\pi e\left( \sigma^2 + N \left( 1 +
%\mathsf{\frac{SNR_2}{1+INR_2}} \right) \right) \right)
%\label{H_Y_2_Y_1_X_1}
%\end{eqnarray}
\begin{eqnarray}
\lefteqn{h(\hat{Y}_2 | Y_1, X_1)} \nonumber \\
&=& \frac{1}{2} \log  \left(2\pi e\left( \sigma^2 + N \left( 1 +
\mathsf{\frac{SNR_2}{1+INR_2}} \right) \right) \right)
\label{H_Y_2_Y_1_X_1}
\end{eqnarray}
where $\sigma^2$ is given by (\ref{sigma_quantize_typeI}).
Now, the rate of user $1$ is given by
\begin{eqnarray}
R_1 &=& I(X_1; Y_1, \hat{Y}_2) \nonumber \\
%& =& I(X_1; Y_1) + I(X_1; \hat{Y}_2 | Y_1) \nonumber \\
& =& I(X_1; Y_1) + h(\hat{Y}_2 | Y_1) - h(\hat{Y}_2 | Y_1, X_1).
\label{R_1_quantization_expression}
\end{eqnarray}

Clearly, with independent Gaussian inputs $X_1 \sim \mathcal{N}(0,
P_1)$ and $X_2 \sim \mathcal{N}(0, P_2)$,
\begin{equation} \label{I_X_1_Y_1}
I(X_1; Y_1) = \gamma \left( \mathsf{\frac{SNR_1}{1+INR_2}} \right).
\end{equation}
Substituting (\ref{I_X_1_Y_1}), (\ref{H_Y_2_Y_1_X_1}) and
$h(\hat{Y}_2|Y_1)$ from (\ref{sigma_and_Y_2_Y_1}) into
(\ref{R_1_quantization_expression}), after some calculations, we
obtain $R_1$ in (\ref{rate_pair_quantization}).

\subsection{Proof of Theorem 3} \label{appendix_theorem_3}

We first prove the achievability of the rate region given in
(\ref{rate_region_moderate_bin_typeII}). We then
show that (\ref{rate_region_moderate_bin_typeII}) reduces to
(\ref{rate_region_weak_typeII}) in the weak interference regime,
and reduces to (\ref{capacity_region_strong_Z_typeII}) and
(\ref{capacity_region_very_strong_Z_typeII}) in the strong and
very strong interference regimes, respectively.

A two-step decoding procedure is used to prove the achievability. Consider first the decoding of $(X_1^n, W_2^n)$ at $Y_1$. The achievable set of $(S_1, T_2)$ is the capacity region of a multiple-access channel, denoted by
$\mathcal{C}_1$, which is just (\ref{end}) with $R_0$ set to zero. Next, consider the decoding of $(W_2^n,U_2^n)$ at receiver $2$ with the help of a digital relay link of rate $R_0$. This is a multiple-access channel with a rate-limited relay, where the relay has complete knowledge of $W_2^n$ and a noisy observation $h_{21}U_2^n + Z_1^n$, obtained by subtracting $X_1^n$ and $W_2^n$ from the received signal at receiver $1$.
Each of these two pieces of information is useful for decoding
$(W_2^n,U_2^n)$ at receiver $2$.

Now, consider a relay scheme which splits the digital link in two parts: $R_a$ bits for describing $U_2^n$, and $R_b$ for describing $W_2$, where $R_a+R_b=R_0$. However, since only a noisy version of $U_2^n$ is available at the relay ($Y_1$), a compress-and-forward strategy using Wyner-Ziv coding
(\cite{WynerZiv, Cover1979}) may be used for describing $U_2^n$.  One
way to do compress-and-forward is to quantize $h_{21} U_2^n + Z_1^n$
with $Y_2^n$ acting as the decoder side information. However, the
presence of $W_2^n$ offers other possibilities. First, receiver $2$
may choose to decode $W_2^n$ before decoding $U_2^n$, in which
case $W_2^n$ becomes additional decoder side information for Wyner-Ziv
coding.  Second, instead of quantizing $h_{21}U_2^n+Z_1^n$ with $W_2^n$
completely subtracted from the relay's observation, the relay may
choose to subtract $W_2^n$ partially---doing so can benefit the
Wyner-Ziv rate. This second approach is is adopted in the rest of the proof. Interestingly, the two approaches turn out to give identical achievable rates.

\begin{figure} [t]
\centering \psfrag{U2}{$U_2$} \psfrag{W2}{$W_2$} \psfrag{Z}{$Z_2 \sim
\mathcal{N}(0, N)$} \psfrag{Y2}{$Y_2$}\psfrag{R0}{$R_0$}
\psfrag{R1}{$R_a$} \psfrag{R2}{$R_b$} \psfrag{h2}{$h_2$}
\psfrag{Y1star}{$\bar{Y}_1$} \psfrag{Y1h}{$\hat{Y}_1$}
\includegraphics[width=1.8in]{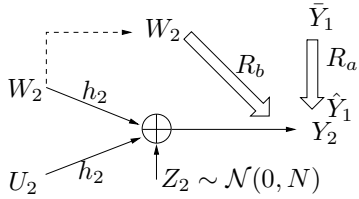}
\caption{Gaussian multiple-access channel with two digital relay
links.} \label{mac_relay_typeii}
\end{figure}

Specifically, let the relay form the following fictitious signal
\begin{equation}
\label{y1_bar_alpha}
\bar{Y}_1^n = h_{21} (U_2^n + W_2^n) + \alpha h_{21} W_2^n + Z_1^n
\end{equation}
for some $\alpha \in \mathbb{R}$. The proposed relay scheme, which
combines the decode-and-forward technique and the compress-and-forward
technique, is illustrated in Fig.~\ref{mac_relay_typeii}, where
$W_2^n$ and $U_2^n$ are the inputs of the multiple-access channel, $(Y_2^n,
\hat{Y}_1^n)$ is the output, and $\hat{Y}_1^n$ is a quantized version of
$\bar{Y}_1^n$. With complete knowledge of $W_2^n$ at the relay, the capacity of
this multiple-access relay channel, denoted by $\mathcal{C}_2$, is
given by the set of rates $(S_2, T_2)$ where
\begin{equation} \label{C2_I_typeII}
\left\{
  \begin{array}{rll}
S_2 &\le& I(U_2; Y_2, \hat{Y}_1 | W_2) \\
T_2 &\le& I(W_2; Y_2, \hat{Y}_1 | U_2) +R_b \\
S_2 + T_2 &\le& I(U_2, W_2; Y_2, \hat{Y}_1) + R_b
  \end{array}
\right.
\end{equation}

Similar to Theorem~\ref{theorem_z_type_I}, we adopt $\bar{Y}_1$: $\hat{Y}_1 = \bar{Y}_1 + e$, where $e$ is a Gaussian random variable independent of $\bar{Y}_1$,
with a distribution $\mathcal{N}(0, \sigma^2)$.  With the encoder
side information $W_2$ at the input of the relay link and the decoder
side information $Y_2$ at the output of the relay link, the Wyner-Ziv
coding rate for quantizing $\bar{Y}_1$ into $\hat{Y}_1$ is given
by (\cite{Chiang} \cite{Cover1979}) $I(\hat{Y}_1; W_2, \bar{Y}_1) - I(\hat{Y}_1; Y_2) \le R_a$.
%\begin{equation}
%I(\hat{Y}_1; W_2, \bar{Y}_1) - I(\hat{Y}_1; Y_2) \le R_a.
%\end{equation}
But
\begin{eqnarray}
\lefteqn{I(\hat{Y}_1; W_2, \bar{Y}_1) - I(\hat{Y}_1; Y_2)} \nonumber \\
&=& I(\hat{Y}_1; \bar{Y}_1) + I(\hat{Y}_1; W_2 | \bar{Y}_1) -
I(\hat{Y}_1; Y_2) \nonumber \\
&\stackrel{(a)}{=}& I(\hat{Y}_1; \bar{Y}_1) - I(\hat{Y}_1; Y_2) \nonumber \\
&\stackrel{(b)}{=}&  I(\hat{Y}_1; \bar{Y}_1 | Y_2)
\end{eqnarray}
where both $(a)$ and $(b)$ come from the fact that
$\hat{Y}_1=\bar{Y}_1 + e$ and $e$ is independent of $W_2$ or $Y_2$.
Thus, we have $I(\hat{Y}_1; \bar{Y}_1 | Y_2) \le R_a$. To fully utilize the channel, we set $\hat{Y}_1$ to be such
that $I(\hat{Y}_1; \bar{Y}_1| Y_2)$ is equal to $R_a$. To find
$\sigma^2$, note that
\begin{equation}
R_a = h(\hat{Y}_1| Y_2) - h(\hat{Y}_1|\bar{Y}_1,Y_2) = \frac{1}{2} \log \left(\frac{\sigma_{\hat{Y}_1|Y_2}^2}{\sigma^2} \right)
\label{sigma_and_Y_1_star}
\end{equation}
where $\sigma_{\hat{Y}_1|Y_2}^2$ is the conditional variance of $\hat{Y}_1$ given $Y_2$. Calculating $\sigma_{\hat{Y}_1|Y_2}^2$ and
substituting it into
(\ref{sigma_and_Y_1_star}), we obtain (\ref{sigma}).

Now, define $I(U_2;\hat{Y}_1|Y_2,W_2)
 \stackrel{\bigtriangleup}{=} \zeta(\alpha,\beta,R_a)$, $I(W_2;\hat{Y}_1|Y_2,U_2) \stackrel{\bigtriangleup}{=} \xi(\alpha,\beta,R_a)$, and $I(W_2,U_2;\hat{Y}_1|Y_2) \stackrel{\bigtriangleup}{=} \eta(\alpha,\beta,R_a)$. Applying Gaussian distributions $W_2 \sim \mathcal{N}(0, \overline{\beta} P_2)$ and $U_2 \sim \mathcal{N}(0, \beta P_2)$, the multiple-access relay channel capacity region
$\mathcal{C}_2$ in (\ref{C2_I_typeII}) becomes
\begin{equation} \label{C2_typeII}
\left\{
  \begin{array}{rll}
S_2 &\le&  \D \gamma (\beta \mathsf{SNR_2}) +
	\zeta(\alpha,\beta,R_a)  \\
T_2 &\le& \gamma (\overline{\beta} \mathsf{SNR_2}) +
	\xi(\alpha,\beta,R_a) + R_b \\
S_2+T_2  &\le& \D  \gamma (\mathsf{SNR_2}) +
	\eta(\alpha,\beta,R_a) + R_b.
  \end{array}
\right.
\end{equation}
The computations of $\zeta(\alpha,\beta,R_a)$,
$\xi(\alpha,\beta,R_a)$ and $\eta(\alpha,\beta,R_a)$
are as follows. First,
\begin{eqnarray}
\eta(\alpha,\beta,R_a) = \frac{1}{2} \log \left(\frac{\sigma_{\hat{Y}_1|Y_2}^2 }{ N+\sigma^2 } \right).
\end{eqnarray}
Calculating $\sigma_{\hat{Y}_1|Y_2}^2$, we obtain
(\ref{delta_12}).  Likewise,
\begin{eqnarray}
\zeta(\alpha,\beta,R_a) = \frac{1}{2} \log \left(\frac{\sigma_{\hat{Y}_1|Y_2,W_2}^2 }
	{ N+\sigma^2 } \right).
\end{eqnarray}
A similar computation leads to (\ref{delta_1}). The expression of
$\xi(\alpha,\beta,R_a)$ does not affect our final result.

Finally, an achievable rate region for the Gaussian Z-relay-interference channel is a set of $(R_1,R_2)$ with $R_1=S_1$ and
$R_2=S_2+T_2$, for which $(S_1,T_2) \in \mathcal{C}_1$ and
$(S_2,T_2) \in \mathcal{C}_2$. Combining the $\mathcal{C}_1$ region and the $\mathcal{C}_2$ region (\ref{C2_typeII}) using the Fourier-Motzkin elimination procedure, we
obtain a pentagon achievable rate region $R_{\alpha,\beta}(R_a,R_b)$
for each fixed $\alpha$, $0 \le \beta \le 1$ and $R_a+R_b=R_0$ as shown in (\ref{R_beta_typeII}).
%\begin{equation}
%\label{R_alpha_beta_typeII}
%\left\{ (R_1, R_2)\left|
%\begin{array}{l}
%R_1 \le \gamma (\mathsf{\frac{SNR_1}{1 + \beta INR_2}}) \\
%R_2 \le \min \left\{ \gamma(\mathsf{SNR_2}) + R_b +
%	\eta(\alpha,\beta,R_a), \right. \\
%	\qquad \qquad \quad \gamma(\mathsf{\beta SNR_2}) + \gamma \left(
%	\mathsf{\frac{\overline{\beta}INR_2}{1+\beta INR_2}} \right) \\
%	\qquad \qquad \quad \left. + \zeta(\alpha,\beta,R_a) \right \} \\
%R_1+ R_2 \le  \gamma(\beta \mathsf{SNR_2}) + \gamma \left(
%	\mathsf{\frac{SNR_1 + \overline{\beta}INR_2}{1 + \beta INR_2}}\right) \\
%	\qquad \qquad \quad + \zeta(\alpha,\beta, R_a)
%\end{array}
%\right. \right \}.
%\end{equation}
With time-sharing, the overall achievable rate region is given by (\ref{rate_region_moderate_bin_typeII}). In the following, we show that
(\ref{rate_region_weak_typeII}),
%(\ref{rate_region_moderate_bin_typeII}),
(\ref{capacity_region_strong_Z_typeII}) and
(\ref{capacity_region_very_strong_Z_typeII}) are all included in the
above achievable rate region.

First, consider the weak interference regime, where $\mathsf{INR_2 \le
SNR_2}$. For any nonnegative $R_b$ and when $\mathsf{INR_2 \le
SNR_2}$, it is easy to verify that
\begin{equation}
\gamma(\mathsf{\beta SNR_2}) + \gamma \left(
\mathsf{\frac{\overline{\beta}INR_2}{1 + \beta INR_2}} \right )\le
\gamma(\mathsf{SNR_2}) + R_b
\end{equation}
and $\zeta(\alpha,\beta, R_a) \le \eta(\alpha,\beta,R_a)$. Thus, the second term of the minimization in the expression of $R_2$ in (\ref{R_beta_typeII}) is always less than the first term.
In this case, $R_a$ enters the rate region
expression only through $\zeta(\alpha,\beta,R_a)$. It is easy to
verify that $\zeta(\alpha,\beta,R_a)$ is a monotonically
increasing function of $R_a$. Thus, the maximum achievability
region is obtained for $R_a=R_0$ and $R_b=0$. Therefore a pure
quantization scheme is optimal in the weak interference regime.

Further, $\alpha$ enters the rate region expression only through
$\zeta(\alpha,\beta,R_0)$. Thus, we can choose $\alpha$ to maximize
$\zeta(\alpha,\beta,R_0)$, or equivalently, to minimize $\sigma^2$
in (\ref{sigma}). Taking the derivative of (\ref{sigma}) on $\alpha$ and setting it to zero, the optimal $\alpha$ is
\begin{equation}
\label{alpha_star}
\alpha^* = - \frac{1}{1+ \beta \mathsf{SNR_2}}.
\end{equation}
Substituting $\alpha^*$ into (\ref{sigma}), we obtain
\begin{equation}
\frac{\sigma^2}{N} = \frac{1}{2^{2R_0}-1}
\left(1+\frac{\beta \mathsf{INR_2}} {1+\beta \mathsf{SNR_2}}\right),
\end{equation}
which gives a derivation of (\ref{delta}):
\begin{equation}
\zeta(\alpha^*,\beta,R_0) =
\gamma \left( \frac{\beta
(2^{2R_0}-1)\mathsf{INR_2}}{2^{2R_0}(\mathsf{1 +\beta SNR_2}) +
\mathsf{\beta INR_2}}\right)
\stackrel{\bigtriangleup}{=}
\delta(\beta, R_0).
\end{equation}

%Finally, we take the union of all $\mathcal{R}_{\alpha^*,\beta}(R_0,0)$.
%Denote the rate constraints
%of the pentagon as
%\begin{eqnarray}
%f_1(\beta) &=& \gamma \left(\mathsf{\frac{SNR_1}
%    {1 + \beta INR_2}} \right) \nonumber \\
%f_2(\beta) &=& \gamma \left(\mathsf{\beta SNR_2}\right) +
%    \gamma \left(\mathsf{\frac{\overline{\beta} INR_2}
%        {1 +\beta INR_2}} \right) + \delta(\beta, R_0) \nonumber \\
%f_3(\beta) &=& \gamma \left(\mathsf{\beta SNR_2}\right) +
%    \gamma \left( \mathsf{\frac{SNR_1 + \overline{\beta}INR_2}
%	{1 + \beta INR_2}} \right) +\delta(\beta,R_0). \nonumber
%\end{eqnarray}

%It is easy to check that, first, the pentagon reduces to a rectangle
%when $\beta=1$. Second, $f_2(\beta)-f_3(\beta)$ is a constant.
%Third, when $\mathsf{INR_2 \le SNR_2}$, we have $f_1^{'}(\beta) < 0$
%and $f_2^{'}(\beta)=f_3^{'}(\beta) \ge 0$. Therefore, as $\beta$
%decreases from 1 to 0, $f_1(\beta)$ is monotonically increasing, while
%both $f_2(\beta)$ and $f_3(\beta)$ are monotonically decreasing.
%This is exactly the same situation as the proof of
%Theorem~\ref{theorem_z_type_I}; see Fig.~\ref{pentagon_union_weak}.

Finally, we take the union of all $\mathcal{R}_{\alpha^*,\beta}(R_0,0)$.
Following the same approach of the proof in Theorem~\ref{theorem_z_type_I},
we can show that the union of achievable pentagons,
$\bigcup_{0 \le \beta \le 1}\mathcal{R}_{\alpha^*,\beta}(R_0,0)$
is defined by $R_1 \le \gamma(\mathsf{SNR_1})$,
$R_2 \le \gamma(\mathsf{SNR_2})+\delta(\beta,R_0)$, and
lower-right corner points of the pentagons
\begin{equation} \label{convex_region_chapter3}
\left\{
  \begin{array}{l}
R_1 = \gamma \left( \D \frac{\mathsf{SNR}_1}{1 + \beta \mathsf{INR}_2}
\right) \\
R_2 = \D \gamma(\beta \mathsf{SNR}_2) + \gamma \left(
	\frac{\overline{\beta} \mathsf{INR}_2}
	{1 + \mathsf{SNR_1} + \beta \mathsf{INR}_2} \right)
	+ \delta(\beta, R_0). \end{array} \right.
\end{equation}
We prove in Appendix \ref{appendix_convexity_type II} that this
region is convex when $\mathsf{INR_2 \le SNR_2}$. Thus, the
convex hull is not needed. This establishes the region
(\ref{rate_region_weak_typeII}) for the weak interference regime.

In the moderately strong interference regime, the achievability of
(\ref{rate_region_moderate_bin_typeII}) follows directly from the
general achievability region. In this regime, the rate region is
achieved by a mixed scheme, which includes both the decode-and-forward
and the compress-and-forward strategies.

Finally, consider the strong interference regime, where
$\mathsf{INR_2 \ge INR_2^{\dagger}}$ and the very strong interference
regime, where $\mathsf{INR_2 \ge INR_2^{\ddag}}$.
We show that (\ref{capacity_region_strong_Z_typeII}) and
(\ref{capacity_region_very_strong_Z_typeII}) are the capacity regions,
respectively.

First, by setting\footnote{The value of $\alpha$ does not affect
$\mathcal{R}_{\alpha,\beta}(R_a,R_b)$ when $R_a=0$.}
$R_b=R_0$, $R_a=0$ and $\beta=0$, the achievable rate region
$\mathcal{R}_{\alpha,\beta}(R_a,R_b)$ in (\ref{R_beta_typeII})
reduces to
\begin{equation} \label{simplified_region_typeII}
\left\{ (R_1, R_2) \left|
  \begin{array}{rll}
R_1 &\le& \gamma (\mathsf{SNR_1})\\
R_2 &\le& \min \left\{  \gamma (\mathsf{SNR_2}) +R_0,
	\gamma (\mathsf{INR_2})  \right\}\\
R_1+R_2 &\le& \gamma(\mathsf{SNR_1 + INR_2})
  \end{array}
\right. \right \}.
\end{equation}
This rate region reduces to (\ref{capacity_region_strong_Z_typeII})
in the strong interference regime, because $\gamma (\mathsf{SNR_2}) +R_0 \le  \gamma (\mathsf{INR_2})$ when $\mathsf{INR_2 \ge INR_2^{\dagger}}$.
Thus, (\ref{capacity_region_strong_Z_typeII}) is achievable.

Further, in the very strong interference regime, where
$\mathsf{INR_2 \ge INR_2^{\ddag}}$, the constraint on $R_1+R_2$ in
(\ref{capacity_region_strong_Z_typeII}) becomes redundant. Thus, the
rate region reduces to (\ref{capacity_region_very_strong_Z_typeII}).

Next, we give a converse proof to show that
(\ref{capacity_region_strong_Z_typeII}) and
(\ref{capacity_region_very_strong_Z_typeII}) are indeed the capacity
regions in the strong and very strong interference regimes,
respectively. Following the same idea as in the converse proof of
Theorem~\ref{theorem_z_type_I}, we show that if $(R_1, R_2)$ is in the
achievable rate region for the Type II channel, i.e., $X_1^n$ can be reliably decoded at receiver $1$ at rate $R_1$, and $X_2^n$ can be reliably decoded at receiver $2$ at rate $R_2$, then $X_2^n$ must also be decodable at the receiver $1$.

First, observe that by the cut-set upper bound \cite{EIT:Cover},
reliable decoding of $X_2^n$ at receiver $2$ requires
\begin{equation} \label{R_2_direct_typeII}
R_2 \le \gamma(\mathsf{SNR_2}) + R_0.
\end{equation}
To show that $X_2^n$ must be decodable at receiver $1$, note that after the
decoding of $X_1^n$ at receiver $1$, $X_1^n$ can be subtracted from the
received signal to form
\begin{equation} \label{R_2_cross_typeII}
\tilde{Y}_1^n =h_{21}X_2^n +Z_1^n.
\end{equation}
The capacity of this channel is $\mathsf{\gamma(INR_2)}$. On the other
hand, $R_2$ is bounded by $\gamma(\mathsf{SNR_2}) + R_0$, which is
less than $\mathsf{\gamma(INR_2)}$ when $\mathsf{INR_2 \ge INR_2^\dagger}$.
So, $X_2^n$ is always decodable based on $\tilde{Y}_1^n$.

Now, since both $X_1^n$ and $X_2^n$ are decodable at receiver $1$ in the strong interference regime, the achievable rate region of the Gaussian Z-relay-interference channel in the strong interference regime must be a subset of the capacity region of a Gaussian multiple-access channel with $X_1^n$, $X_2^n$ as inputs
and $Y_1^n$ as output, which is
\begin{equation} \label{sss}
\left\{ (R_1, R_2)\left|
  \begin{array}{rll}
R_1 &\le& \mathsf{\gamma(SNR_1)}  \\
R_2 &\le& \mathsf{\gamma(INR_2)} \\
R_1 +R_2 &\le& \mathsf{\gamma(SNR_1 +INR_2)}
  \end{array}
\right. \right \}.
\end{equation}
Combining (\ref{R_2_direct_typeII}), (\ref{sss}), and observing that $\gamma (\mathsf{SNR_2}) +R_0 \le  \gamma (\mathsf{INR_2})$ when $\mathsf{INR_2 \ge INR_2^{\dagger}}$, we proved that the
achievable rate region of the Gaussian Z-relay-interference channel
must be bounded by (\ref{capacity_region_strong_Z_typeII}) when
$\mathsf{INR_2 \ge INR_2^{\dagger}}$,
which reduces to (\ref{capacity_region_very_strong_Z_typeII}) when
$\mathsf{INR_2 \ge INR_2^{\ddag}}$.

%The achievability proof for the weak interference case in
%Theorem~\ref{theorem_Z_typeII} is based on a compress-and-forward
%scheme in which $Y_1$ quantizes a fictitious signal
%$\bar{Y}_1=h_{21}(U_2+W_2) + \alpha^* h_{21} W_2 + Z_1$, which is a
%linear combination of its own received signal and the decoded $W_2$,
%with $\alpha^*$ {\em optimized} for maximum overall achievable rate
%region.  The quantized $\bar{Y}_1$ is decoded with
%$Y_2=h_{22}(U_2+W_2)+Z_2$ as decoder side information in establishing
%the multiple-access relay channel capacity region $\mathcal{C}_2$ with
%Wyner-Ziv coding.
%
%As mentioned earlier, another possible compress-and-forward strategy
%is to restrict the decoding order for the multiple-access channel
%$\mathcal{C}_2$ to be that of decoding $W_2$ first, then $U_2$. In
%this case, $W_2$ would be known at {\em both} the input and the output
%of the relay link when decoding $U_2$. Thus, the relay only has to
%quantize $h_{21}U_2+Z_1$ with $h_{22}U_2+Z_2$ as decoder side
%information in Wyner-Ziv coding.
%
%Interestingly, these two different strategies give the exact same
%achievable rate region in the weak-interference regime. This fact is
%shown in Appendix \ref{app_U_only}, which gives an alternative proof
%for the weak-interference result in Theorem~\ref{theorem_Z_typeII}.

\subsection{Convexity of Achievable Rate Region (\ref{convex_region_chapter3})}
\label{appendix_convexity_type II}

This appendix proves that the region defined by $R_1 \le
\mathsf{\gamma(SNR_1)}$, $R_2 \le \mathsf{\gamma(SNR_2)}+
	\delta(\beta, R_0)$, and the curve
\begin{equation} \label{concave_curve_typeII}
\left\{
\begin{array}{lll}
R_1 & \le & \gamma \left ( \D \frac{\mathsf{SNR}_1}{1 +
    \beta \mathsf{INR}_2} \right) \\
R_2 & \le & \D  \gamma(\beta \mathsf{SNR}_2) +
    \gamma \left( \frac{\overline{\beta}\mathsf{INR}_2}{1 +
    \mathsf{SNR}_1 + \beta \mathsf{INR}_2} \right) +\delta(\beta, R_0)
\end{array}
\right.
\end{equation}
where $0 \le \beta \le 1$, is convex when $\mathsf{INR_2 \le
SNR_2}$.

We follow the same idea used in Appendix \ref{appendix_convexity}
to prove the convexity of the above region.
By Appendix \ref{appendix_convexity}, we can rewrite $R_2$ as
\begin{equation} \label{R_2_as_functin_of_R_1}
R_2 = \frac{1}{2} \log \left( -\nu 2^{2R_1} + \lambda \right) +
\tilde{\mu} + \delta(\beta, R_0)
\end{equation}
where $\tilde{\mu} = \mu - R_0$ is a constant, and $\nu, \lambda,
\mu$ are as defined in Appendix \ref{appendix_convexity}.

%Since $\delta_1(\beta, R_0)$ is a monotonic increasing function of
%$\beta$, and $\beta$ as a function of $R_1$ (see
%(\ref{beta_as_function_of_R1})) is a monotonic decreasing function
%of $R_1$, we conclude that, $\delta_1(\beta, R_0)$ as a function of
%$R_1$, is monotonic decreasing. Therefore, $R_2$, which is the sum
%of $\delta_1(\beta, R_0)$ and a decreasing term $\frac{1}{2} \log
%\left( -\nu 2^{2R_1} + \lambda \right) + \tilde{\mu}$, is also a
%monotonic decreasing function of $R_1$. Thus
%\begin{equation}
%R_2' \le 0
%\end{equation}

It is easy to verify that in the weak interference regime,
$\delta(\beta, R_0)$ is concave in $\beta$, and
$\beta(R_1)$, as denoted in (\ref{beta_as_function_of_R1}), is
convex in $R_1$. Combining this with the fact that $\delta(\beta,
R_0)$ is a nondecreasing function of $\beta$ shows that
$\delta(\beta, R_0)$ is a concave function of $R_1$. Adding
$\delta(\beta, R_0)$ with another concave (proved in Appendix
\ref{appendix_convexity}) term $\frac{1}{2} \log \left( -\nu
2^{2R_1} + \lambda \right) + \tilde{\mu}$ gives us the desired result
that $R_2$ is a concave function of $R_1$.

Therefore, the region defined by $R_1 \le \mathsf{\gamma(SNR_1)}$,
$R_2 \le \mathsf{\gamma(SNR_2)}+\delta(\beta, R_0)$ and
(\ref{concave_curve_typeII}) is convex.

\bibliographystyle{IEEEtran}
%\bibliography{IEEEabrv,./ref/main}

\begin{IEEEbiographynophoto}{Lei Zhou}
(S'05) received the B.E. degree in electronics engineering
from Tsinghua University, Beijing, China,
in 2003 and M.A.Sc. degree in electrical and computer engineering from the University of Toronto, ON, Canada, in 2008. During 2008-2009, he was with Nortel Networks, Ottawa, ON, Canada. He is currently pursuing the Ph.D. degree with the Department of Electrical and Computer Engineering, University of Toronto, Canada. His research interests include multiterminal information theory, wireless communications, and signal processing.

He is a recipient of the Shahid U.H. Qureshi Memorial Scholarship in 2011, and the Alexander Graham Bell Canada Graduate Scholarship for 2011-2013.
\end{IEEEbiographynophoto}

\begin{IEEEbiographynophoto}{Wei Yu}
(S'97-M'02-SM'08) received the B.A.Sc. degree in Computer Engineering and
Mathematics from the University of Waterloo, Waterloo, Ontario, Canada in 1997
and M.S. and Ph.D. degrees in Electrical Engineering from Stanford University,
Stanford, CA, in 1998 and 2002, respectively. Since 2002, he has been with the
Electrical and Computer Engineering Department at the University of Toronto,
Toronto, Ontario, Canada, where he is now an Associate Professor and holds a
Canada Research Chair in Information Theory and Digital Communications. His
main research interests include multiuser information theory, optimization,
wireless communications and broadband access networks.

Prof. Wei Yu currently serves as an Associate Editor for
{\sc IEEE Transactions on
Information Theory} and an Editor for {\sc IEEE Transactions on Communications}.
He was an Editor for {\sc IEEE Transactions on Wireless
Communications} from 2004 to 2007, and a Guest Editor for a number of
special issues for the {\sc IEEE Journal on Selected Areas in
Communications} and the {\sc EURASIP Journal on Applied Signal Processing}.
He is member of the Signal Processing for Communications and Networking
Technical Committee of the IEEE Signal Processing Society.
He received the IEEE Signal Processing Society Best Paper Award in 2008,
the McCharles Prize for Early Career Research Distinction in 2008,
the Early Career Teaching Award from the Faculty of Applied Science
and Engineering, University of Toronto in 2007, and the Early Researcher
Award from Ontario in 2006.
\end{IEEEbiographynophoto}

\end{document}